\documentstyle[10pt,epsf,epsfig,hangcaption,xspace,amssymb,amsfonts,amsmath,amsthm,cite,
dp_delphititle,lineno]{dp_delphi}
\makeindex
\def\DpPaperGroup{EP}
\def\DpPaperRef{2003-065}
\def\DpDate{28 Mai 2003}
\def\DpAuthors{DELPHI Collaboration}
\def\DpSubmit{(Accepted by Eur. Phys. J.)}
\def\DpTitle{{A Precise Measurement of the \\
 \BP, \BZERO~and Mean b-hadron\\
 Lifetime with the DELPHI \\
Detector at LEP I }}
\def\DpComment{ }
\def\DpEMail{ }

\newcommand{\BZERO} {${\mathrm B^0}$}
\newcommand{\BP} {${\mathrm B^+}$}
\newcommand{\BS} {${\mathrm B}_{{\mathrm s}}^0$}
\newcommand{\BD} {${\mathrm B}_{{\mathrm d}}^0$}
\newcommand{\Ztobb} {${\mathrm Z}^0 \rightarrow {\mathrm b}\bar{{\mathrm b}}$}
\newcommand{\Ztoqq} {${\mathrm Z}^0 \rightarrow {\mathrm q}\bar{{\mathrm q}}$}
\newcommand{\LB} {${\mathrm \Lambda_b}$}

\newcommand{\micron} {$\mu {\mathrm m}$}
\newcommand{\meanb} {$\tau_{\mathrm b} $}
\newcommand{\vcb} {$\left| V_{{\mathrm cb}} \right|$}
\begin{document}
\makeatletter
\newcount\@tempcntc
\def\@citex[#1]#2{\if@filesw\immediate\write\@auxout{\string\citation{#2}}\fi
  \@tempcnta\z@\@tempcntb\m@ne\def\@citea{}\@cite{\@for\@citeb:=#2\do
    {\@ifundefined
       {b@\@citeb}{\@citeo\@tempcntb\m@ne\@citea\def\@citea{,}{\bf ?}\@warning
       {Citation `\@citeb' on page \thepage \space undefined}}%
    {\setbox\z@\hbox{\global\@tempcntc0\csname b@\@citeb\endcsname\relax}%
     \ifnum\@tempcntc=\z@ \@citeo\@tempcntb\m@ne
       \@citea\def\@citea{,}\hbox{\csname b@\@citeb\endcsname}%
     \else
      \advance\@tempcntb\@ne
      \ifnum\@tempcntb=\@tempcntc
      \else\advance\@tempcntb\m@ne\@citeo
      \@tempcnta\@tempcntc\@tempcntb\@tempcntc\fi\fi}}\@citeo}{#1}}
\def\@citeo{\ifnum\@tempcnta>\@tempcntb\else\@citea\def\@citea{,}%
  \ifnum\@tempcnta=\@tempcntb\the\@tempcnta\else
   {\advance\@tempcnta\@ne\ifnum\@tempcnta=\@tempcntb \else \def\@citea{--}\fi
    \advance\@tempcnta\m@ne\the\@tempcnta\@citea\the\@tempcntb}\fi\fi}

\makeatother
\begin{titlepage}
\pagenumbering{roman}
\CERNpreprint{\DpPaperGroup}{\DpPaperRef} 
\date{{\small\DpDate}} 
\title{\DpTitle} 
\address{\DpAuthors} 
\begin{shortabs} 
\noindent
%
\noindent

Final results from the DELPHI Collaboration
on the  lifetime of \BP~and \BZERO~mesons and the mean b-hadron lifetime,
are presented using
the data collected at the ${\mathrm Z}^0$~peak in 1994 and 1995.
Elaborate, inclusive, secondary vertexing methods have been
employed to ensure a b-hadron
reconstruction with good efficiency. To separate samples of
\BP~and \BZERO~mesons, high performance neural network techniques are
used that achieve very high purity signals.
The results obtained are:
\begin{tabbing}
tttttt\=ttt\=tttttt\=tttttttttttttttt\=ttttttttttttt \kill
\hspace{2.5cm}
$\tau_{{\mathrm B}^+}$ \>\hspace{2.5cm} = \>\hspace{2.5cm} 1.624\>\hspace{2.5cm} $\pm 0.014$~(stat)  \>\hspace{2.5cm}$\pm 0.018 $~(syst)~ps \\  

\hspace{2.65cm}$\tau_{{\mathrm B}^0}$ \>\hspace{2.5cm} = \>\hspace{2.5cm} 1.531 \>\hspace{2.5cm} $\pm 0.021$~(stat)  \>\hspace{2.5cm}$\pm 0.031 $~(syst)~ps \\

\hspace{2.45cm} $\frac{\tau_{{\mathrm B}^+}}{\tau_{{\mathrm B}^0}}$ \>\hspace{2.5cm} = \> \hspace{2.5cm} 1.060 \>\hspace{2.6cm}$\pm 0.021$~(stat)\> \hspace{2.5cm}$\pm 0.024$~(syst)
\end{tabbing}
and for the average b-hadron lifetime:
\begin{tabbing}
tttttt\=ttt\=tttttt\=tttttttttttttttt\=ttttttttttttt \kill
\hspace{2.65cm}$\tau_{{\mathrm b}}$ \>\hspace{2.5cm} = \>\hspace{2.5cm} 1.570 \>\hspace{2.5cm} $\pm 0.005$~(stat)  \>\hspace{2.5cm}$\pm 0.008$~(syst)~ps.
\end{tabbing}
\end{shortabs}
\vfill
\begin{center}
\DpSubmit \ \\ 
\DpComment \ \\
\DpEMail \ \\
\end{center}
\vfill
\clearpage
\headsep 10.0pt
\addtolength{\textheight}{10mm}
\addtolength{\footskip}{-5mm}
\begingroup
%
\newcommand{\DpName}[2]{\hbox{#1$^{\ref{#2}}$},\hfill}
\newcommand{\DpNameTwo}[3]{\hbox{#1$^{\ref{#2},\ref{#3}}$},\hfill}
\newcommand{\DpNameThree}[4]{\hbox{#1$^{\ref{#2},\ref{#3},\ref{#4}}$},\hfill}
\newskip\Bigfill \Bigfill = 0pt plus 1000fill
\newcommand{\DpNameLast}[2]{\hbox{#1$^{\ref{#2}}$}\hspace{\Bigfill}}
%
\footnotesize
\noindent
\DpName{J.Abdallah}{LPNHE}
\DpName{P.Abreu}{LIP}
\DpName{W.Adam}{VIENNA}
\DpName{P.Adzic}{DEMOKRITOS}
\DpName{T.Albrecht}{KARLSRUHE}
\DpName{T.Alderweireld}{AIM}
\DpName{R.Alemany-Fernandez}{CERN}
\DpName{T.Allmendinger}{KARLSRUHE}
\DpName{P.P.Allport}{LIVERPOOL}
\DpName{U.Amaldi}{MILANO2}
% ...
\normalsize
\endgroup
\titlefoot{Department of Physics and Astronomy, Iowa State
     University, Ames IA 50011-3160, USA
    \label{AMES}}
\titlefoot{Physics Department, Universiteit Antwerpen,
     Universiteitsplein 1, B-2610 Antwerpen, Belgium \\
     \indent~~and IIHE, ULB-VUB,
     Pleinlaan 2, B-1050 Brussels, Belgium \\
     \indent~~and Facult\'e des Sciences,
     Univ. de l'Etat Mons, Av. Maistriau 19, B-7000 Mons, Belgium
    \label{AIM}}
\titlefoot{Physics Laboratory, University of Athens, Solonos Str.
     104, GR-10680 Athens, Greece
    \label{ATHENS}}
\titlefoot{Department of Physics, University of Bergen,
     All\'egaten 55, NO-5007 Bergen, Norway
    \label{BERGEN}}
\titlefoot{Dipartimento di Fisica, Universit\`a di Bologna and INFN,
     Via Irnerio 46, IT-40126 Bologna, Italy
    \label{BOLOGNA}}
\titlefoot{Centro Brasileiro de Pesquisas F\'{\i}sicas, rua Xavier Sigaud 150,
     BR-22290 Rio de Janeiro, Brazil \\
     \indent~~and Depto. de F\'{\i}sica, Pont. Univ. Cat\'olica,
     C.P. 38071 BR-22453 Rio de Janeiro, Brazil \\
     \indent~~and Inst. de F\'{\i}sica, Univ. Estadual do Rio de Janeiro,
     rua S\~{a}o Francisco Xavier 524, Rio de Janeiro, Brazil
    \label{BRASIL}}
\titlefoot{Coll\`ege de France, Lab. de Physique Corpusculaire, IN2P3-CNRS,
     FR-75231 Paris Cedex 05, France
    \label{CDF}}
\titlefoot{CERN, CH-1211 Geneva 23, Switzerland
    \label{CERN}}
\titlefoot{Institut de Recherches Subatomiques, IN2P3 - CNRS/ULP - BP20,
     FR-67037 Strasbourg Cedex, France
    \label{CRN}}
\titlefoot{Now at DESY-Zeuthen, Platanenallee 6, D-15735 Zeuthen, Germany
    \label{DESY}}
\titlefoot{Institute of Nuclear Physics, N.C.S.R. Demokritos,
     P.O. Box 60228, GR-15310 Athens, Greece
    \label{DEMOKRITOS}}
\titlefoot{FZU, Inst. of Phys. of the C.A.S. High Energy Physics Division,
     Na Slovance 2, CZ-180 40, Praha 8, Czech Republic
    \label{FZU}}
\titlefoot{Dipartimento di Fisica, Universit\`a di Genova and INFN,
     Via Dodecaneso 33, IT-16146 Genova, Italy
    \label{GENOVA}}
\titlefoot{Institut des Sciences Nucl\'eaires, IN2P3-CNRS, Universit\'e
     de Grenoble 1, FR-38026 Grenoble Cedex, France
    \label{GRENOBLE}}
\titlefoot{Helsinki Institute of Physics, P.O. Box 64,
     FIN-00014 University of Helsinki, Finland
    \label{HELSINKI}}
\titlefoot{Joint Institute for Nuclear Research, Dubna, Head Post
     Office, P.O. Box 79, RU-101 000 Moscow, Russian Federation
    \label{JINR}}
\titlefoot{Institut f\"ur Experimentelle Kernphysik,
     Universit\"at Karlsruhe, Postfach 6980, DE-76128 Karlsruhe,
     Germany
    \label{KARLSRUHE}}
\titlefoot{Institute of Nuclear Physics,Ul. Kawiory 26a,
     PL-30055 Krakow, Poland
    \label{KRAKOW1}}
\titlefoot{Faculty of Physics and Nuclear Techniques, University of Mining
     and Metallurgy, PL-30055 Krakow, Poland
    \label{KRAKOW2}}
\titlefoot{Universit\'e de Paris-Sud, Lab. de l'Acc\'el\'erateur
     Lin\'eaire, IN2P3-CNRS, B\^{a}t. 200, FR-91405 Orsay Cedex, France
    \label{LAL}}
\titlefoot{School of Physics and Chemistry, University of Lancaster,
     Lancaster LA1 4YB, UK
    \label{LANCASTER}}
\titlefoot{LIP, IST, FCUL - Av. Elias Garcia, 14-$1^{o}$,
     PT-1000 Lisboa Codex, Portugal
    \label{LIP}}
\titlefoot{Department of Physics, University of Liverpool, P.O.
     Box 147, Liverpool L69 3BX, UK
    \label{LIVERPOOL}}
\titlefoot{Dept. of Physics and Astronomy, Kelvin Building,
     University of Glasgow, Glasgow G12 8QQ
    \label{GLASGOW}}
\titlefoot{LPNHE, IN2P3-CNRS, Univ.~Paris VI et VII, Tour 33 (RdC),
     4 place Jussieu, FR-75252 Paris Cedex 05, France
    \label{LPNHE}}
\titlefoot{Department of Physics, University of Lund,
     S\"olvegatan 14, SE-223 63 Lund, Sweden
    \label{LUND}}
\titlefoot{Universit\'e Claude Bernard de Lyon, IPNL, IN2P3-CNRS,
     FR-69622 Villeurbanne Cedex, France
    \label{LYON}}
\titlefoot{Dipartimento di Fisica, Universit\`a di Milano and INFN-MILANO,
     Via Celoria 16, IT-20133 Milan, Italy
    \label{MILANO}}
\titlefoot{Dipartimento di Fisica, Univ. di Milano-Bicocca and
     INFN-MILANO, Piazza della Scienza 2, IT-20126 Milan, Italy
    \label{MILANO2}}
\titlefoot{IPNP of MFF, Charles Univ., Areal MFF,
     V Holesovickach 2, CZ-180 00, Praha 8, Czech Republic
    \label{NC}}
\titlefoot{NIKHEF, Postbus 41882, NL-1009 DB
     Amsterdam, The Netherlands
    \label{NIKHEF}}
\titlefoot{National Technical University, Physics Department,
     Zografou Campus, GR-15773 Athens, Greece
    \label{NTU-ATHENS}}
\titlefoot{Physics Department, University of Oslo, Blindern,
     NO-0316 Oslo, Norway
    \label{OSLO}}
\titlefoot{Dpto. Fisica, Univ. Oviedo, Avda. Calvo Sotelo
     s/n, ES-33007 Oviedo, Spain
    \label{OVIEDO}}
\titlefoot{Department of Physics, University of Oxford,
     Keble Road, Oxford OX1 3RH, UK
    \label{OXFORD}}
\titlefoot{Dipartimento di Fisica, Universit\`a di Padova and
     INFN, Via Marzolo 8, IT-35131 Padua, Italy
    \label{PADOVA}}
\titlefoot{Rutherford Appleton Laboratory, Chilton, Didcot
     OX11 OQX, UK
    \label{RAL}}
\titlefoot{Dipartimento di Fisica, Universit\`a di Roma II and
     INFN, Tor Vergata, IT-00173 Rome, Italy
    \label{ROMA2}}
\titlefoot{Dipartimento di Fisica, Universit\`a di Roma III and
     INFN, Via della Vasca Navale 84, IT-00146 Rome, Italy
    \label{ROMA3}}
\titlefoot{DAPNIA/Service de Physique des Particules,
     CEA-Saclay, FR-91191 Gif-sur-Yvette Cedex, France
    \label{SACLAY}}
\titlefoot{Instituto de Fisica de Cantabria (CSIC-UC), Avda.
     los Castros s/n, ES-39006 Santander, Spain
    \label{SANTANDER}}
\titlefoot{Inst. for High Energy Physics, Serpukov
     P.O. Box 35, Protvino, (Moscow Region), Russian Federation
    \label{SERPUKHOV}}
\titlefoot{J. Stefan Institute, Jamova 39, SI-1000 Ljubljana, Slovenia
     and Laboratory for Astroparticle Physics,\\
     \indent~~Nova Gorica Polytechnic, Kostanjeviska 16a, SI-5000 Nova Gorica, Slovenia, \\
     \indent~~and Department of Physics, University of Ljubljana,
     SI-1000 Ljubljana, Slovenia
    \label{SLOVENIJA}}
\titlefoot{Fysikum, Stockholm University,
     Box 6730, SE-113 85 Stockholm, Sweden
    \label{STOCKHOLM}}
\titlefoot{Dipartimento di Fisica Sperimentale, Universit\`a di
     Torino and INFN, Via P. Giuria 1, IT-10125 Turin, Italy
    \label{TORINO}}
\titlefoot{INFN,Sezione di Torino, and Dipartimento di Fisica Teorica,
     Universit\`a di Torino, Via P. Giuria 1,\\
     \indent~~IT-10125 Turin, Italy
    \label{TORINOTH}}
\titlefoot{Dipartimento di Fisica, Universit\`a di Trieste and
     INFN, Via A. Valerio 2, IT-34127 Trieste, Italy \\
     \indent~~and Istituto di Fisica, Universit\`a di Udine,
     IT-33100 Udine, Italy
    \label{TU}}
\titlefoot{Univ. Federal do Rio de Janeiro, C.P. 68528
     Cidade Univ., Ilha do Fund\~ao
     BR-21945-970 Rio de Janeiro, Brazil
    \label{UFRJ}}
\titlefoot{Department of Radiation Sciences, University of
     Uppsala, P.O. Box 535, SE-751 21 Uppsala, Sweden
    \label{UPPSALA}}
\titlefoot{IFIC, Valencia-CSIC, and D.F.A.M.N., U. de Valencia,
     Avda. Dr. Moliner 50, ES-46100 Burjassot (Valencia), Spain
    \label{VALENCIA}}
\titlefoot{Institut f\"ur Hochenergiephysik, \"Osterr. Akad.
     d. Wissensch., Nikolsdorfergasse 18, AT-1050 Vienna, Austria
    \label{VIENNA}}
\titlefoot{Inst. Nuclear Studies and University of Warsaw, Ul.
     Hoza 69, PL-00681 Warsaw, Poland
    \label{WARSZAWA}}
\titlefoot{Fachbereich Physik, University of Wuppertal, Postfach
     100 127, DE-42097 Wuppertal, Germany \\
\noindent
{$^\dagger$~deceased}
    \label{WUPPERTAL}}
\addtolength{\textheight}{-10mm}
\addtolength{\footskip}{5mm}
\clearpage
\headsep 30.0pt
\end{titlepage}
%
\pagenumbering{arabic} 
\setcounter{footnote}{0} %
\large
\section{Motivation and Overview}
\label{intro}

In addition to testing models of b-hadron
decay,
knowledge of b-hadron lifetimes is of importance in the determination
of other Standard Model quantities such as the CKM matrix element 
$V_{{\mathrm cb}}$
and in measurements of the time dependence of 
neutral b-meson oscillations.

The Spectator Model provides the simplest description of 
b-hadron decay.
Here, the lifetime depends only on the weak decay of the b-quark
with the other light-quark
constituent(s) playing no role in the decay dynamics. This, in turn,
leads to the prediction that all b-hadron species have the 
same lifetime. 
Non-spectator effects however such as quark interference,
$\mathrm{W}$~exchange and weak annihilation can induce lifetime
differences among the different b-hadron species.
Models of b-hadron decay based on expansions in $1/m_{{\mathrm b}}$
predict that, 
$\tau({\mathrm \Lambda_b}) < \tau({\mathrm B^0})
      \sim \tau({\mathrm B_s}) < \tau({\mathrm B^+})$,\footnote {Note that 
the corresponding charge conjugate state is always implied throughout 
this paper. \BZERO~always refers to the ${\mathrm B_d^0}$~state
 and $\Lambda_b$~refers to any, weakly decaying, b-baryon.} 
and this lifetime hierarchy has already 
been confirmed by experiment.
There is a growing consensus between models that a difference 
in lifetime of order 5\% should exist between the 
\BP~and \BZERO~meson \cite{THEORY1} and it is 
currently measured to be,  
$\frac{\tau_{{\mathrm B}^+}}{\tau_{{\mathrm B}^0}}=1.085 \pm 0.017$~\cite{PDG}.
Clearly, more precise lifetime measurements of all b-hadron species are valuable in order
to test developments in b-hadron decay theory. 


This paper reports on the measurement of the \BP~and \BZERO~meson lifetimes
from data taken with the DELPHI detector at LEP I, in sub-samples separately  enriched
in \BP~and \BZERO~mesons. 
In addition, 
a measurement of the mean b-hadron lifetime \meanb~(i.e. with
$ {\mathrm \Lambda_b},{\mathrm B^0},{\mathrm B_s},{\mathrm B^+}$ unseparated) 
is obtained, which is a quantity of importance for many 
b-physics analyses at LEP e.g. in the extraction of the CKM matrix 
element \vcb.     

The analysis proceeds by reconstructing the proper time of b-hadron candidates
($t=Lm_0/pc$~where $L,p$~and $m_0$ are the decay length, momentum and 
rest mass respectively) and fitting the $t$~distribution in simulation
to the data distribution in a $\chi^2$--minimisation procedure.
In the fit for the  \BP~and \BZERO~lifetimes, the ${\mathrm B_s}$~and
${\mathrm \Lambda_b}$~lifetimes are set to their measured values
whereas in the fit for \meanb, all B-species are assumed to have the same
lifetime and \meanb~is the only free parameter in the fit.
The approach used is highly inclusive
and based  on
the DELPHI inclusive B-physics package, BSAURUS \cite{BSAURUS}.  
Aspects of BSAURUS directly related to the analysis are 
presented in a summarised form but the reference should 
be consulted for full details of the package. 

After describing parts of the DELPHI detector essential for this
measurement in Section \ref{sec:DELPHI}, the data sets are described
in Section~\ref{sec:gensel} and   
relevant aspects of BSAURUS are highlighted in Section~\ref{sec:bsaurus}. 
Section~\ref{sec:propt} describes the reconstruction of the B-candidate proper
time from measurements of decay length and momentum. 
Samples with a $\sim 70$\% purity in \BP~or \BZERO~mesons
were achieved by use of a sophisticated neural network approach 
which is described in Section~\ref{sec:enhance}.
 Section~\ref{sec:results} shows the results of the lifetime fits and
finally, systematic uncertainties on the measurements
are dealt with in Section \ref{sec:syst}.

\section{The DELPHI Detector}
\label{sec:DELPHI}

A complete overview of the  DELPHI detector \cite{DELPHI} and of its 
performance \cite{DELPHI_PERF} have been 
given in detail elsewhere. 
This analysis depends crucially on precision charged particle tracking
performed by the  
Vertex Detector (VD), the Inner Detector, the Time Projection Chamber
(TPC) and the Outer Detector. A highly uniform magnetic field of
1.23 T parallel to the electron beam direction, was provided by the 
superconducting solenoid throughout the tracking volume. The momentum 
of charged particles was measured with a precision
of  $\sigma_p/p \leq 1.5\%$~in the polar angle region $40^\circ \le \theta \le 140$~and 
for $p<10$~GeV/$c$.

The VD was of particular importance for the reconstruction
of the decay vertices of short-lived particles and consisted of
three layers of  silicon
microstrip detectors, called the Closer, Inner and Outer layers,
 at radii of 6.3\,cm, 9\,cm and 11\,cm
from the beam line respectively. The VD was upgraded \cite{NEWVD}
in 1994 and 1995 
with double-sided microstrip detectors in the Closer and Outer layers
providing coordinates\footnote{DELPHI has a 
cylindrical polar coordinate system. The $z$-axis points along the 
beam direction and $r$~and $\phi$~are the radial and azimuthal coordinates
in the transverse plane.} 
in both $r\phi$ and $rz$. 
For polar angles of $44^\circ \le \theta \le 136^\circ$, a track crosses
all three silicon layers of the VD. 
The measured intrinsic resolution is about 8 \micron~for the $r\phi$
coordinate
while for $rz$ it depends on the incident polar angle of the track and reaches
about 9 \micron~for tracks perpendicular to the modules.
For  tracks with hits in all three $r\phi$ VD layers
the impact parameter  resolution was
$\sigma_{r\phi}^2=([61 / (p\, \sin ^{3/2} \theta)]^2+ 20^2)$~\micron$^2$~and  
for tracks with hits in both $rz$ layers and with $\theta \approx 90^\circ$,
$\sigma_{rz}^2=([67 / (p\, \sin ^{5/2} \theta )]^2+ 33^2)$~\micron$^2$.       
Before the start of data taking in 1995 the ID was replaced with a similar
device but with a larger polar
angle coverage in preparation for LEP 2 running. The impact of this change on
the current analysis is relatively minor.
 
Calorimeters detected photons and neutral hadrons by the total 
absorption of their energy. The High-density Projection
Chamber (HPC) provided electromagnetic calorimetry coverage in the 
polar angle region $46^{\circ}<\theta<134^{\circ}$~giving a 
relative precision on the measured energy $E$~of 
$\sigma_E/E=0.32/\sqrt{E} \oplus 0.043$~($E$~in GeV). 
In addition, each HPC module is essentially a small TPC
which can chart the spatial development of showers and
so provide an angular resolution better than that of the
detector granularity alone. For high energy photons
the angular resolutions were $\pm 1.7$~mrad in the azimuthal
angle $\phi$~and  $\pm 1.0$~mrad in the polar angle
$\theta$.

The
Hadron Calorimeter was installed in the return yoke of the 
DELPHI solenoid and provided a relative precision on the 
measured energy of
$\sigma_E/E=1.12/\sqrt{E} \oplus 0.21$~($E$~in GeV).
 
Powerful particle identification (see Section~\ref{sec:partid}) 
was possible by the combination of
$dE/dx$~information from the TPC (and to
a lesser extent from the VD) with information from
the DELPHI  Ring Imaging CHerenkov counters (RICH) in 
both the forward and barrel regions. The RICH devices
utilised both liquid and gas radiators in order to 
optimise coverage across a wide momentum range:
liquid was used for the momentum range from
0.7 GeV/$c$ to 8 GeV/$c$ and the gas radiator for the 
range 2.5 GeV/$c$ to 25 GeV/$c$. 


\section{Data Samples}
\label{sec:gensel}

\subsection{Hadronic Event Selection}
Hadronic ${\mathrm Z^0}$ decays were selected by the following requirements:
\begin{itemize}
\item at least 5 reconstructed charged particles,
\item the summed energy of charged particles (with momentum $> 0.2$~GeV/$c$)
      had to be larger than 12\% of the centre-of-mass 
      energy, with at least 3\% of it in each of the forward and backward
      hemispheres defined with respect to the beam axis.  
\end{itemize}
Due to the evolution of the DELPHI tracking detectors with time,
and run-specific details such as the RICH efficiency,
the data were treated throughout this analysis as two independent  
data sets for the periods 1994 and 1995. 
The hadronic event cuts selected approximately
1.4 million events in 1994 and 0.7 million events in 1995. 


\subsection{Event Selection} 
\label{SEC:EVSEL}
\begin{figure}[htb]
\begin{center}
\leavevmode
\includegraphics[width=10.0cm]{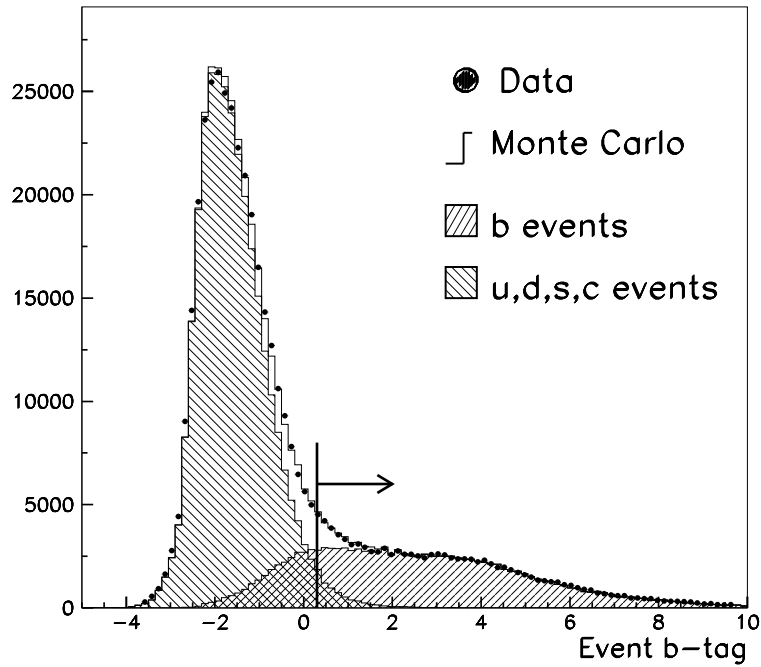}
\caption[]
{\label{FIG:BTAG} \it The event b-tag variable in 1994 data and simulation,
normalised to the number of entries.}
\end{center}
\end{figure}

Hadronic events were enhanced in \Ztobb~events, 
by cutting on the DELPHI combined  b-tagging variable described in 
\cite{AABTAG}. In the construction of the b-tag, the following 
four variables were 
selected that are highly correlated with the presence of a b-hadron 
but only weakly correlated between themselves:
\begin{itemize}
\item[1.] The mass of particles included in  a reconstructed b-hadron secondary vertex.
\item[2.] The probability that if a track originated from
the primary vertex, it  would have a positive
lifetime-signed impact parameter, with respect to this vertex,
at least as large as that observed.
\item[3.] The  fraction of the total jet energy contained in
the tracks  associated with a secondary vertex, fitted with an
algorithm run only on the tracks associated with the jet.  
\item[4.] The rapidity, 
with respect to the jet axis direction (see Section~\ref{sec:rapy}), of tracks 
included in the secondary vertex.
\end{itemize}
These variables were combined in likelihood ratios 
(which assumes they are independent), to give a single b-tag variable
that can be applied to tag single jets, hemispheres
or the whole event. The event b-tag, comparing data and simulation, is shown in 
Figure~\ref{FIG:BTAG} together with the contributions of b- and u,d,s,c-events 
from the simulation. The arrow indicates the position of the cut made to 
enhance the sample in  \Ztobb~events. 
In addition to the b-tag requirement it was demanded that events be 
well contained in the barrel of the DELPHI detector by making the 
following cut on the event thrust axis vector: $\left| \cos(\theta_{thrust}) \right|<0.65$.
After applying the event selection cuts the purity in 
\Ztobb~events was about $\sim 94\%$~and
the data samples
consisted of 285088(136825) event hemispheres in 1994(1995) respectively.

\subsection{Simulated Data}
Simulated data sets were produced with the JETSET 7.3 \cite{JETSET} package
with tunings optimized to DELPHI
data \cite{MCTUNE} and passed through a full detector simulation
\cite{DELPHI_PERF}. Both
\Ztoqq~and \Ztobb~events were used 
and separate samples were produced for comparison with 1994 and 1995 
data. The same hadronic and event selection criteria  were 
applied to the simulation samples as for the data resulting in 
640888(239061) events of ${\mathrm q} \bar{{\mathrm q}}$~Monte Carlo in 
1994(1995)    
and 1581499(422304) events of ${\mathrm b}\bar{{\mathrm b}}$~Monte Carlo. 

\section{General Tools}
\label{sec:bsaurus} 

The identification of b-hadron candidates and the reconstruction of their
decay length and momentum (or equivalently their energy), was made in 
a completely inclusive way; i.e. the analysis was sensitive to all
 b-hadron decay channels. 
This section describes briefly some tools that were essential in being able
to decide whether the data were likely to contain a b-hadron and 
what the probability was, for that b-hadron, to be a \BP,\BZERO,\BS~or 
b-baryon. Two specially constructed neural networks (the TrackNet and the
BD-Net, described below) made it possible to distinguish whether a particle was likely to 
have originated from the primary vertex (in the fragmentation process or the decay 
of an excited state), from the weakly decaying b-hadron secondary vertex or even
from the 
${\mathrm B} \rightarrow {\mathrm D}$~cascade tertiary vertex. 
The reconstruction 
of the b-hadron secondary vertex position was essential in order to reconstruct the 
decay length, which in turn tags the presence of a long-lived b-hadron state. 
Exploiting the excellent particle identification ability of the DELPHI detector 
was another aid to b-hadron reconstruction. The presence of high-momentum electrons or
muons is a powerful signature of a  b-hadron decay and the detection of kaons or protons
also provides valuable information on the type of b-hadron.

\subsection{Rapidity}
\label{sec:rapy}
Events were split into two hemispheres using the plane perpendicular to the thrust axis. 
A reference axis was defined in each hemisphere along a jet
 reconstructed 
via the routine LUCLUS \cite{JETSET}~with $p_{\bot}$~as a distance measure and
the cutoff parameter $d_{join}$(PARU(44))=5.0 GeV/$c$. In simulation studies this
was found to give the best reconstruction of the initial b-hadron
direction. For hemispheres where two or more jets were reconstructed (about
16\% of the cases), the b-tag applied at the jet level was
used to discriminate the b-jet from the gluon jet. With this scheme, the
probability to select correctly the two b-jets in a three-jet event was 
about 70\%.

\begin{figure}[htb]
\begin{center}
\leavevmode
\includegraphics[width=8.0cm]{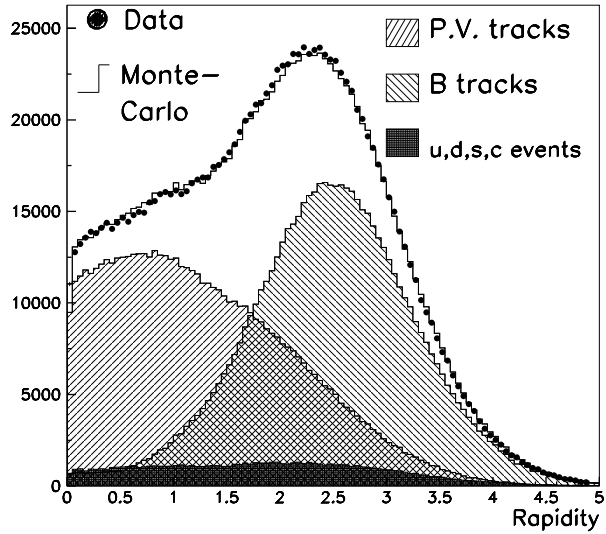}
\caption[]
{\label{FIG:RAPIDITY} \it Distributions of rapidity from 1994 data and simulation
where the normalisation is to the number of entries. From the simulation,
tracks originating at the primary vertex and tracks from the b-hadron decay chain 
are also plotted.} 
\end{center}
\end{figure}

The rapidity, with respect to the reference axis, was defined as
\begin{eqnarray*}
 y & = &\frac{1}{2} \cdot \ln \left((E+P_{\|})/(E-P_{\|})\right),
\end{eqnarray*}
where identified particles
were assigned their respective masses and all others were
assigned a pion mass.
Figure~\ref{FIG:RAPIDITY}, shows the rapidity distribution for real and
simulated data after the event selection cuts have been applied. 
Particles originating from the decay chain of a b-hadron are seen 
to have higher mean rapidities than  particles originating from the 
primary vertex. This property made the rapidity a useful 
input  quantity to the  TrackNet (Section~\ref{sec:tracknet})
and for 
b-hadron decay vertex reconstruction (Section~\ref{sec:secver}).
In addition, a {\it Rapidity Algorithm} was defined that 
summed $(\vec{p}_i,E_i,m_{i})$~for particles with  $y>1.6$~in 
order to form an estimate of the weakly decaying b-hadron four-vector. 
\subsection{The TrackNet}
\label{sec:tracknet} 
The TrackNet is the output of a neural network which
supplied, for each track in the hemisphere, 
the probability that the track originates from the decay of the
b-hadron.
The network 
relied on the presence of a reconstructed secondary vertex (see Section~\ref{sec:secver})
providing an estimate of the b-hadron decay position
 in the event hemisphere.
The b-hadron four-vector was reconstructed by the Rapidity
Algorithm described above. 
The most important inputs to the 
neural network were:  
\begin{itemize}
\item the magnitude of the particle momentum in the laboratory frame,
\item the magnitude of the particle momentum in the B-candidate rest frame,
\item the track  helicity angle, defined as the angle between 
the track vector in the B-candidate rest frame and the B-candidate momentum vector
in the lab frame,
\item a flag to identify whether the track was in the secondary vertex
      fit or not,
\item the primary vertex track probability (defined as in Point 2. of Section~\ref{SEC:EVSEL}),
\item the secondary vertex track probability (defined as in Point 2. of Section~\ref{SEC:EVSEL}),
\item the particle rapidity.
\end{itemize} 

\begin{figure}[h]
\begin{center}
\leavevmode
\includegraphics[width=10.0cm]{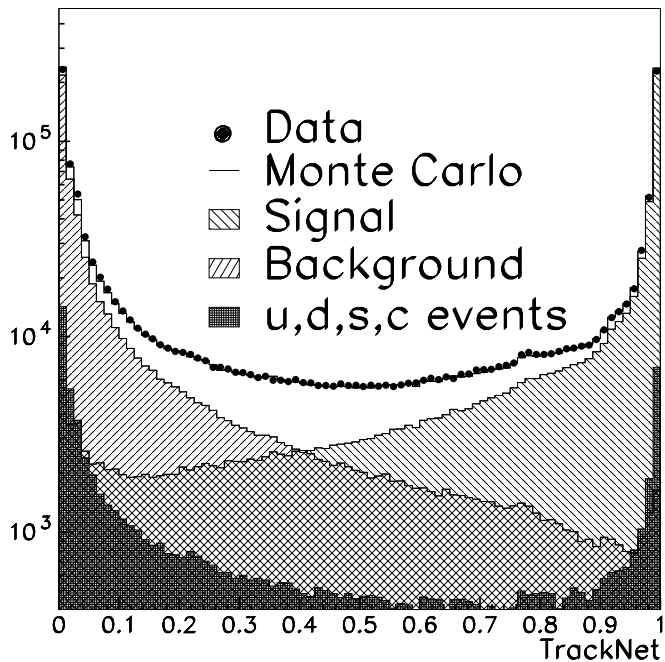}
\caption[]
{\label{fig:tnet} \it Distributions of the TrackNet from 1994 data and simulation, 
normalised  to the number of entries.} 
\end{center}
\end{figure}

Figure~\ref{fig:tnet}, shows the TrackNet distribution for real and
simulated data after the event selection cuts have been applied. 
The simulation distribution is divided into contributions from
\Ztobb~events representing $\sim 94\%$~of the 
total and the remaining $6\%$, shown in black, is due to tracks in 
u,d,s-quark and c-quark  ${\mathrm Z^0}$~decays. 
The b-events are further divided into `signal' and `background'.
The signal consisted of tracks
originating from the b-hadron weak decay chain and this class of 
track formed the target for the neural network training. The background 
consisted of everything else in b-events such as fragmentation tracks and 
decay products of excited b-hadrons. Figure~\ref{fig:tnet} illustrates 
the excellent separation, of b-hadron decay  products from other tracks,
 possible using
the TrackNet and shows a good overall agreement in shape between data and
simulation.
As a result of the b-tagging cut, the contribution from u,d,s,c-events is
dominated by c-events ($\sim 80\%$) which 
account for the spike around  TrackNet$=1.0$~in the u,d,s,c distribution. 

\subsection{Secondary Vertex Finding}
\label{sec:secver}
For each hemisphere an attempt was made to fit a secondary vertex 
from charged  particle tracks 
that were likely to
have originated from the decay chain of a weakly decaying b-hadron.
Tracks were first selected by imposing the following set of standard quality cuts:
\begin{itemize}
\item impact parameter in the $r\phi$ plane $\left| \delta_{r-\phi} \right| < 4.0$~cm,
\item impact parameter in the $z$ plane $\left| \delta_{z} \right| < 6.0$~cm,
\item $\left| \cos \theta \right| < 0.94$,
\item $\frac{\Delta E}{E} < 1.0$,
\item at least one $r-\phi$ track hit from the silicon vertex detector(VD),
\item tracks must not have been flagged as originating from interactions with
  detector material by the standard DELPHI interaction vertex reconstruction
  package, described in \cite{SIGMA}.
\end{itemize}
In addition, all charged particles were required to have 
rapidity greater than the value 1.6, giving good discrimination
between tracks orginating from the fragmentation process and those 
originating from the b-hadron decay chain (see Figure~\ref{FIG:RAPIDITY}). 
In order to be considered further for vertex 
fitting, charged particles were selected using a procedure based on:
\begin{itemize}
\item The likelihood to be an electron, muon or kaon.
\item The `lifetime' content of the track based on the 
three-dimensional crossing point of the track with the 
estimated b-hadron direction vector from the Rapidity
Algorithm. 
The direction vector
was forced to pass through  the event primary vertex position.
\item Particle rapidity.
\end{itemize}
If at least two tracks are selected, these tracks are fitted to 
a common secondary vertex
in three dimensions with the constraint that the 
secondary vertex momentum vector must point back  to the
primary vertex error ellipse.
If the fit did not converge\footnote{Non-convergence 
means:
the $\chi^2$ was above 4 standard deviations at the end of the 
first 10 iterations or above 3 standard deviations at the end of
the next 10 iterations or took more than 20 iterations in total.} 
the track making the largest
$\chi^2$~contribution was stripped away, in an iterative
procedure, and the fit repeated.
In a final step, tracks with a large TrackNet value, which had
not already been selected by the initial track search,
were added to the vertex track list and the vertex re-fitted. 

The secondary vertex described above 
refers to the `standard' vertex fit which
was an essential input to the   
calculation of other b-physics quantities used in the analysis
e.g. the BD-Net described in Section~\ref{sec:bdnet}. More sophisticated
secondary vertexing algorithms developed specifically for  the 
extraction of B lifetimes are described in Section~\ref{sec:decl}.

\subsection{Particle Identification}
\label{sec:partid}
The MACRIB package \cite{MACRIB}
provided separate neural networks for the tagging of kaons and protons 
which combined the various sources of particle identification 
in DELPHI.
An efficiency for the correct identification of 
$K^{\pm}$~of 90\%(70\%) was attained with a contamination of 
15\%(30\%) for $p<0.7$ GeV/$c$ ($p>0.7$ GeV/$c$). The corresponding 
contamination  for proton identification
at the same efficiencies was, 2\%(40\%) for  $p<0.7$ GeV/$c$ ($p>0.7$ GeV/$c$). 

Electron and muon candidates were defined according to the standard DELPHI
lepton identification criteria.
Only muon and electron candidates with energy larger than 3 GeV were selected. 

\subsection{The BD-Net}
\label{sec:bdnet}
In selecting tracks for inclusion in the B secondary vertex fit,
there is inevitably some 
background from tracks that originate not from the B decay vertex
directly, but from the subsequent D cascade decay. When such tracks
are present the vertex is in general reconstructed somewhere in the 
region between the B decay point and the D vertex and the 
reconstructed decay length will be biased to larger values.
The resolution of the B decay vertex is therefore improved if 
these tracks can be identified and removed from the secondary vertex.  
In order to identify these tracks a neural network (the BD-Net)
was developed based on the following discriminating variables:
\begin{itemize}  
\item the angle between the track vector and the reconstructed  B flight 
direction, 
\item the primary vertex track probability (defined as in Point 2. of Section~\ref{SEC:EVSEL}),
\item the secondary vertex track probability (defined as in Point 2. of Section~\ref{SEC:EVSEL}),
\item the momentum and angle of the track vector in the B rest frame,
\item the TrackNet output defined in Section~\ref{sec:tracknet},
\item the kaon network output, described in Section~\ref{sec:partid},
\item the lepton identification tag, mentioned in Section~\ref{sec:partid}.
\end{itemize}

\begin{figure}[h]
\begin{center}
\leavevmode
\includegraphics[width=10.0cm]{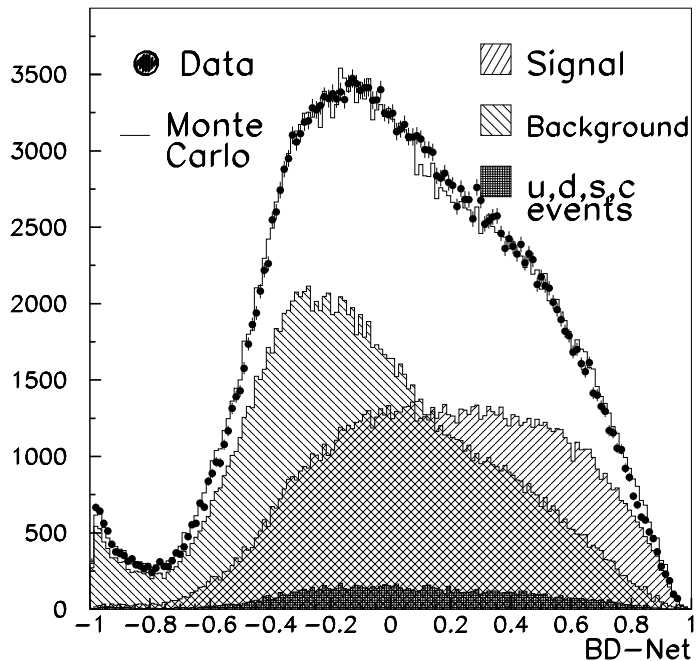}
\caption[]
{\label{fig:bdnet} \it Distributions of the BD-Net from 1994 data and simulation,
normalised to the number of entries.}
\end{center}
\end{figure}

The network was trained to recognise 
tracks originating from  the 
decay chain ${\mathrm B \rightarrow  D \rightarrow X}$~(`signal') compared to 
all other tracks in b-events (`background') where, in addition, all 
tracks must have TrackNet values larger than $0.5$.    
Figure~\ref{fig:bdnet} shows the BD-Net variable for data 
and simulation after the event selection cuts have been applied
plus the same track selection cuts 
that are used in the vertex finding algorithms 
(see Section~\ref{sec:decl}).
The normalisation is to the number of entries and the 
simulation has been weighted to adjust the 
b-hadron production fractions, the B$\rightarrow$D branching ratios and 
the `wrong-sign' 
\footnote{Charm quarks produced 
from the upper- or W-vertex in b-hadron decay via e.g. 
${\mathrm W}^+ \rightarrow {\mathrm c} {\mathrm \bar{s}}$.} 
${\mathrm D_s}$~charm production rate to the same values as 
detailed in Section~\ref{sec:systa}.
The discrimination attained between signal and background is also
plotted together with the distribution shape expected from 
tracks in u,d,s,c-events. The background distribution in 
Figure~\ref{fig:bdnet} is dominated by tracks from weak b-hadron decays
 and the 
small spike at values close to BD-Net$=-1$~is due to semi-leptonic decays 
of b-hadrons which are readily recognised by the network as not
coming from the D-vertex. The agreement between data and simulation
is  good and the effect of any residual discrepancy on the 
analysis was found to be insignificant (see Section~\ref{sec:systa}). 
 

%

\section{Proper Decay Time Reconstruction}
\label{sec:propt}
This section deals with the reconstruction of the
proper time defined as, 

\begin{eqnarray}
\nonumber
t=Lm_0/pc
\end{eqnarray}
where $L$~and $p$~are the reconstructed decay length and momentum of the
B-candidate  respectively and $m_0$~is the B rest mass which was taken to be
 $5.2789$ GeV$/c^2$.
The magnitude of the B-candidate momentum vector was fixed by the
relationship $E^2=p^2+m_0^2$, where $E$~is the reconstructed B-candidate energy. 
The reconstruction of the two essential components of the proper time, namely 
the B-candidate energy and the decay length, are now dealt with in  
some detail.

\subsection{B-candidate Energy Reconstruction}
\label{sec:benergy}
A novel method to reconstruct the  B-candidate energy was used which involved
training a neural network (EB-Net) to return a complete  probability 
density function (p.d.f.) for the energy on a {\it hemisphere-by-hemisphere} basis.
Training the network proceeds by 
dividing up the inclusive truth B-energy distribution from the Monte Carlo
$f(B)$~into many equal slices or threshold levels.
One neural network output node is assigned to each level and 
the target value for each training event becomes a vector with as many elements as there are 
output nodes, defined by the classification: `is the energy of the B in 
this hemisphere, higher (1) or 
lower (-1) than the associated threshold at this output node'. 
The median of  the resulting estimate of $f(B)$~is taken
as the EB-Net estimator for the B-energy.  


  For the network training sixteen input variables were used which
  included different estimators of the energy available in the hemisphere 
  together with some measures of the expected quality of such estimators
  e.g. as given by such quantities as  hemisphere track multiplicity and
  hemisphere reconstructed energy. The key inputs to the network were:
\begin{itemize}
\item In 2-jet events the 
sum (over all particles $i$~in a hemisphere) of  the  vectors $(\vec{p}_i,E_i,m_{i})$~was 
formed, weighted by the TrackNet value (if particle $i$~was a charged particle)
or  weighted by a function of the rapidity (if particle $i$~was a
neutral cluster).
 In this way 
particles from the b-hadron decay received a higher weight in the sum and 
hence an estimate of the b-hadron vector was obtained: ${\bf P_B}=\left( \vec{p}_B,E_B,m_B \right)$.
 For  $\geq$~3-jet events  ${\bf P_B}$~was estimated via the Rapidity Algorithm.
These estimates of $E_B$~were then used directly as an input to the EB-Net and 
the momentum vectors $\vec{p}_B$~provided the direction constraint  
for the secondary vertex fitting algorithms described in Section~\ref{sec:decl}.

\item   The estimate of the b-hadron energy $E_B$~was corrected to 
account for sources of missing energy. 
The correction procedure was motivated by the observation in simulation
of a correlation between the energy residuals $\Delta E=E_B-E_B^{generated}$
and $m_B$, which is approximately linear in $m_B$, 
and a further correlation between $\Delta E$ and 
$x_h$~(the fraction of the beam  energy in the hemisphere)  resulting
from neutral energy losses and inefficiencies. 
The correction was implemented  
by dividing the data into several samples according to the measured 
ratio $x_h$ and for each of these classes the B-energy residual
$\Delta E$ was formed as function of $m_B$.
The median values of $\Delta E$ in each bin of $m_B$ were calculated   
and their $m_B$-dependence fitted by a third order polynomial
\begin{displaymath}
\Delta E(m_B,x_h)=a+b(m_B-\left<m_B\right>)+
c(m_B-\left<m_B\right>)^2+d(m_B-\left<m_B\right>)^3    
\end{displaymath}
The four parameters ${a,b,c,d}$ in each $x_h$ class were then plotted as a 
function of $x_h$ and their dependence fitted with second and third-order
polynomials. Thus a smooth correction function was obtained from the 
simulation describing the 
dependence of $\Delta E$~on $m_B$ and on the hemisphere energy.
The corrected   b-hadron energy was the 
most powerful input variable to the EB-Net, 
 having a correlation of $73\%$~to the true b-hadron energy.
\end{itemize}

\begin{figure}[tb]
\begin{center}
\leavevmode
\epsfig{file=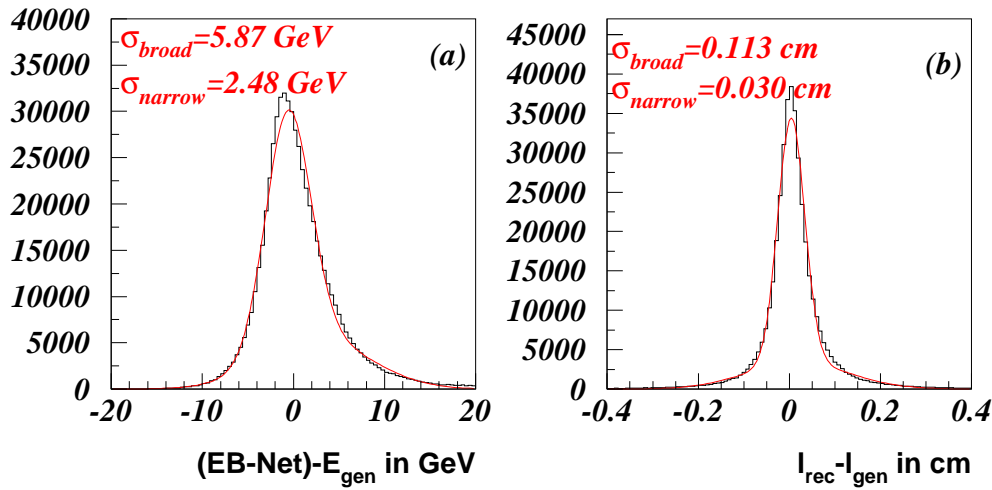,width=14.0cm}
\caption[]
        {\it\label{fig:ereso} 
(a) The EB-Net and (b) the reconstructed B-candidate 
decay length residual i.e. the difference between
the  reconstructed  value and the generated value based 
on 1994 simulated data. }
\end{center}
\end{figure}

The performance of the EB-Net estimator is shown in 
Figure~\ref{fig:ereso}(a) which plots the residual of the 
EB-Net variable with the generated energy value. 
A double-Gaussian fit to the distribution gives a 
central, narrow, Gaussian
covering 67\% of the total area with standard deviation of $2.5$~GeV.
(Note that the fits are approximate and are only to gauge the widths of
the distributions. They are not used in the lifetime measurements.)

\subsection{Decay Length Reconstruction}
\label{sec:decl}

Starting from the standard secondary vertex described in Section~\ref{sec:secver}, four 
algorithms were implemented, based on the BD-Net,
 with the aim of improving the decay length
resolution and minimising any 
bias of the type described in Section~\ref{sec:bdnet}, 
resulting from the inclusion of tracks from the cascade D-decay vertex in the
B-decay vertex reconstruction. In addition to passing the standard quality cuts 
listed in Section~\ref{sec:secver}, tracks were required to have TrackNet values $>0.5$~to
be considered for any of the four algorithms.   
\begin{itemize}
\item [1)] In the {\bf Strip-Down} method candidate tracks were selected if, in 
addition to the cuts described above, they had BD-Net values $<0.0$.
A secondary vertex fit was made if there were two or more tracks selected.
If the fit failed to converge (under the same criteria as were applied to 
the standard fit - see Section~\ref{sec:secver}),
 and more than two tracks were originally selected,
the track with the highest $\chi^2$ contribution was removed and the fit repeated.
 This procedure continued iteratively until convergence was reached or 
less than two tracks were left. 
Technically, the fit was the same as that used to fit  
the standard secondary vertex described in Section~\ref{sec:secver} except 
the starting point was the secondary vertex 
position estimate  of the standard fit. 
\item [2)] In the {\bf D-Rejection} method, a cascade D-candidate vertex
was first built by fitting a common vertex to the two tracks with
the largest BD-Net values in the hemisphere.
If the invariant mass of the combination was below
the D-meson mass, an attempt was made to include also the
track with the next largest BD-Net value. This process continued iteratively
until either the mass exceeded the D-mass, 
there were no further tracks, or the fit failed to converge.
The B-candidate vertex was then fitted using the Strip-Down algorithm but 
applied to all tracks {\it except} those already selected for the D-vertex.   
\item [3)] In the {\bf Build-Up} method those two tracks with TrackNet bigger 
than $0.5$ and smallest BD-Net values
were chosen to form a seed vertex.
If the invariant mass of all remaining tracks with TrackNet~$>0.5$ exceeded the D-mass, 
that track with the lowest BD-Net output was also fitted to a common vertex
with the two seed tracks. 
This process continued iteratively until either the fit failed to converge
or the mass in remaining tracks dropped below the D-meson mass.
\item [4)] The {\bf Semileptonic} algorithm was designed to improve the vertex
  resolution for semileptonic decays of b-hadrons where
  energy has been carried away by the associated neutrino. 
When there was a clear lepton candidate in the hemisphere,
the algorithm reconstructed a cascade D-candidate vertex in a similar way to
the D-rejection method but with the lepton track excluded. The tracks associated
with the vertex were then combined to form a `D-candidate track' which was 
extrapolated back to make the  B-candidate vertex with the lepton track
if the opening angle between the lepton and D-candidate satisfied 
$|\cos \Theta_{LD}|<0.99$.

\end{itemize}

The choice of decay length for the decay time calculation 
was dictated by optimising the resolution and minimising any bias
while still retaining the best possible efficiency. 
If more than one of the four algorithms was successful
in reconstructing a vertex, the choice was 
made in the following order:
\begin{itemize}
\item[1)] the Strip-Down method  was chosen if the algorithm 
had a decay length error smaller than 1 mm,
\item[2)] if the Strip-Down method criteria were not met, the D-rejection method was used
if the decay length error was smaller than 1 mm,
\item[3)] if the criteria for 1) and 2) were not met
 the Build-Up vertex was chosen if the decay length
error was smaller than 200~\micron,
\item[4)]  if the criteria for 1), 2) and 3) were not met
the Semileptonic algorithm was used if the decay length
error was smaller than 1 mm.
\end{itemize}
About one third of all hemispheres, passing
the event selection cuts, were rejected by the 
decay length selection procedure in data and
in simulation.
There were 180010 vertices selected in the 1994
data set and 86796 in 1995. 
Figure~\ref{fig:ereso}(b) plots the residual between
the reconstructed decay length and the generated value.
A double-Gaussian fit to the distribution gives a 
central, narrow, Gaussian
covering 71\% of the total area with a standard deviation of 300~\micron.
The lack of a significant  positive bias to this distribution, illustrates
that the influence of tracks from cascade D-meson decays has been 
successfully  minimised
by employing algorithms based on the BD-Net. 

\section{Selection of \BP~and \BZERO~Enhanced Samples}
\label{sec:enhance}
The enrichment of \BZERO~and \BP~mesons was part of a general attempt
 to provide a probability for an event hemisphere to contain a 
b-hadron of a particular type.
The result was implemented in a neural network ($NN(B_x)$)
consisting of 16 input variables  and a 
4-node output layer.
Each output node delivered a probability for the hypothesis it was trained on: 
the first supplied the probability for \BS~mesons to be produced in the hemisphere, 
the second for \BZERO~mesons, the third for charged B-mesons and the fourth for all 
species of b-baryons. 
The method relied heavily on the reconstruction of the following
quantities:
\begin{itemize}

\item {\bf b-hadron type probabilities $P({\mathrm B}_x)$}: supplied 
by an auxiliary neural network constructed to supply inputs to the
more optimal $NN(B_x)$~network. 
In common with the $NN(B_x)$, there were four output nodes
trained to return the probability that the decaying b-hadron state was 
\BP, \BZERO, \BS~or b-baryon. There were fifteen input
variables in total, the most powerful of which were the hemisphere
TrackNet-weighted charge sum, which 
discriminates charged from neutral states, and variables that exploit
the presence of particular particles produced in association with   
b-hadron states.
Examples of this include \BS~mesons, which are normally produced with a charged kaon as
the leading fragmentation particle with a further kaon emerging from the weak
decay, and in \BP~and \BZERO~production where the decay is associated with a 
larger multiplicity of charged pions than that for \BS~and b-baryons
(which on average will produce a higher proportion of neutrons, protons and kaons).

\item {\bf The b-hadron flavour} i.e. the charge of the 
constituent b-quark
both at the fragmentation ($F_{frag.}$) and decay time ($F_{dec.}$): 
knowledge of b-hadron flavour provides the network with valuable information 
about whether a \BZERO~state was present 
since the fragmentation and decay flavour will,  on average,
disagree for the case where the  \BZERO~oscillated. 
The flavour was determined by first
constructing, with neural network techniques, the conditional probability
for each charged particle in the hemisphere to have the same charge as the b-quark
in the b-hadron. This was repeated separately for each of the 
four possible b-hadron type scenarios i.e. \BZERO,\BP,\BS~or b-baryon. 
The flavour network was trained on a target value of $+1(-1)$~if the 
particle charge was
correlated(anti-correlated) to the b-quark charge. The main input variables 
were those related to the identification of kaons, protons, electrons 
and muons together 
with quantities sensitive to the B-D vertex separation in the hemisphere. 
Tracks originating from the fragmentation
(decay) phase were discriminated by checking that the TrackNet value is
less (greater) than 0.5.
In a final step, these track level probabilities were
combined via a likelihood ratio into  hemisphere quantities.   
Providing separate flavour networks for the different b-hadron 
types not only ensured that the information was optimal for the case of
\BZERO~but also
helped the performance of the enrichment $NN(B_x)$~network by providing 
information that was specific to a particular b-hadron type.

\end{itemize}

The inputs to the $NN(B_x)$~were constructed to exploit optimally all of the
information that the b-hadron production and decay process reveals.
The basic construct for input 
variables was the following  
combination of the flavour and b-hadron type information described above:
\begin{eqnarray}
F_{dec.}({\mathrm B}_x) \cdot F_{frag.}({\mathrm B}_x) \cdot P({\mathrm B}_x).
\end{eqnarray} 

\begin{figure}[tb]
\begin{center}
\leavevmode
\epsfig{file=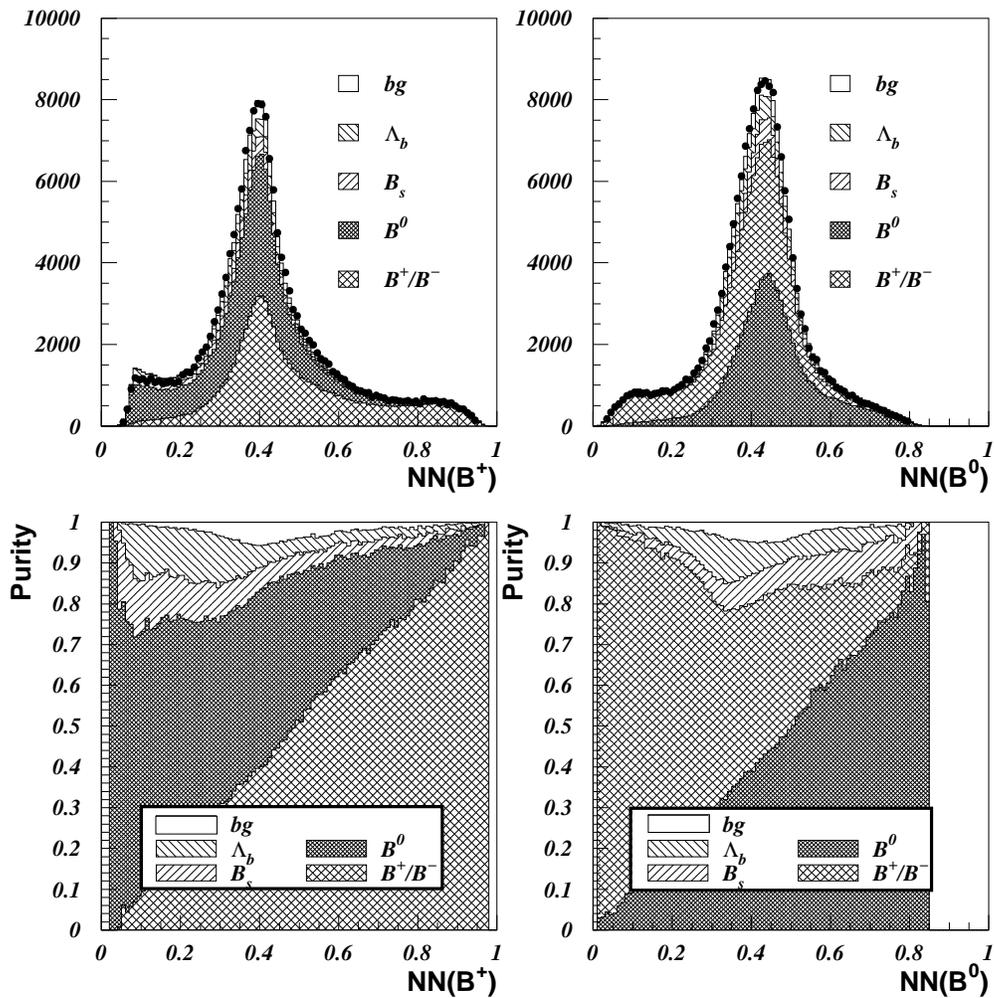,width=13.1cm}
\caption[]
        { \it\label{fig:BHBN} The upper plots show the 
output of the \BP~and \BZERO~output nodes of the $NN(B_x)$~in the 1994 data
and simulation for the different b-hadron types.
The normalisation is
to the number of  data events and 
overlaid is the b-hadron composition as seen in the Monte Carlo.
The lower plots trace the change in purity of the 
 different b-hadron types, per bin, as a function of cuts on  
the $NN(B^+)$~and $NN(B^0)$~respectively.
}
\end{center}
\end{figure}

The upper plots of 
Figure~\ref{fig:BHBN} show the output of the \BP~and \BZERO~output nodes of the $NN(B_x)$
in simulation and data. The simulation is further divided into
the  different b-hadron types and 
the lower plots trace the change in purity of the different types 
in each bin of 
the network output at the \BP~and \BZERO~output nodes respectively. 
Note that the background from u,d,s- and c-events is labelled as  `bg'.

\section{Extraction of \BP~and \BZERO~Lifetimes}
\label{sec:results}
This section describes how the data and simulation samples 
were prepared, gives details of the fitting procedure itself and summarises
the results obtained. Section~\ref{sec:weight} lists corrections made to the
default simulation to account for known discrepancies with data and to 
update b-physics parameters to agree with  recent world
measurements. In  Section~\ref{sec:workpt} the final selection
and composition of the  \BP, \BZERO~and \meanb~samples are described
together with an explanation of how
the region at low proper times i.e. $< 1$~ps was handled in the lifetime
fits. Section~\ref{sec:fit} gives technical details of the $\chi^2$~fit and 
presents the results obtained. Lifetimes were measured separately in 1994 and 1995 data
and then combined to give the final results which are presented in Section~\ref{sec:conclude}. 

\subsection{Simulation Weighting}
\label{sec:weight}
Weights were applied to the simulation 
to correct  for the following effects: 
\begin{itemize}
\item 
The world average of measurements of 
\BS, \LB~lifetimes and b-hadron production fractions,
as compiled for the Winter conferences of  2001~\cite{MORIOND_2001}, 
listed in  Table~\ref{tab:bfracs}. Note that using more ~\cite{MORIOND_2001}
recent world average values has a negligible effect on the
analysis results. 
 
\item The Peterson function used in the  Monte Carlo
($\langle x \rangle=0.706$)
was changed to agree with the functional form  unfolded from 
the 1994 DELPHI data set~\cite{delphi_bfrag} 
($\langle x \rangle=0.7153$).


\item 
  A hemisphere `quality flag' which was proportional to the number of
 tracks in the hemisphere likely to be badly reconstructed e.g. those
 tracks failing the standard selection criteria of Section~\ref{sec:secver}.
The weight was constructed to account for data/simulation discrepancies
in this variable and was formed in bins  of the 
number of charged particles in the hemisphere, that passed the standard quality
cuts of Section~\ref{sec:secver},  
thus ensuring that, overall, the multiplicity of 
 good charged particles was 
unchanged after  the application of the weight.
\end{itemize}
{\small
\begin{table}[h,t,b]
\begin{tabular}{|l|c|c|c|c|} \hline
{\bf b-hadron species} & {\bf Lifetime} &\multicolumn{3}{c|} {{\bf Production fractions $f(b \rightarrow B_x)$}}           \\  \cline{3-5}
                 &           &   {\bf Values}                              & $\rho(B_x,B_s)$  &   $\rho(B_x,baryon)$ \\ \hline
\BS  & $1.464 \pm 0.057$ps &$f({\mathrm B}^0_{\mathrm s})=0.097 \pm 0.011$ &  -               & +0.034             \\ \hline
b-baryons & $1.208 \pm 0.051$ps & $f({\mathrm baryon})=0.104 \pm 0.017$    & $+0.034$         & -                  \\ \hline
\BD~or \BP  & -     & $f({\mathrm B}^0_{\mathrm d})=f({\mathrm B^+})=0.399 \pm 0.010$ & -0.577 & -0.836 \\ \hline
\end{tabular}
\caption[] {\label{tab:bfracs}\it Values for the b-hadron lifetimes and 
production fractions (together with correlations) used to re-weight the  
Monte Carlo. }
\end{table}
}

\subsection{Fit working point}
\label{sec:workpt}

The selection conditions imposed on the data samples used for the lifetime
fits were motivated by the wish to minimise the total error on the final
results. Systematic error contributions due to inexact detector resolution
simulation and the physics modelling of u,d,s and charm production, imply that 
relatively high b-hadron purities were required while still keeping the   
selection efficiency above a level where the statistical error would begin to 
degrade significantly. 

With these considerations in mind, the final data samples
to be used in the fitting procedure were selected 
by cutting 
on the $NN(B_x)$~neural network outputs, described in
  Section~\ref{sec:enhance}, at
$>0.52$~and $>0.6$~respectively to obtain
enhanced samples in \BP~and \BZERO. These cut values corresponded
to a purity in both \BP~and \BZERO~of approximately $70\%$ according to the simulation.
No reconstructed hemisphere passed both the 
\BP~and \BZERO~enhancement cuts simultaneously and hence the two samples 
were statistically independent.

The region below about $1.0$~ps in proper lifetime is 
particularly challenging to simulate. 
The modelling of very small lifetimes is 
rather sensitive to details of reconstruction resolution  and 
the modelling of events which  contain no intrinsic 
lifetime information such as  u,d,s events and the reconstruction
or spurious vertices. In addition, 
the lower plots of Figures~\ref{fig:fits} 
and~\ref{fig:meanfit} show that the purities of the different
b-hadron types is rapidly changing in this region, making them 
particularly difficult to model. These issues meant that the 
low lifetime region was not well enough under control 
systematically for precision lifetime information to be extracted
and the region below  $1.0$~ps was therefore excluded from the analysis.
This point is illustrated 
in Figure~\ref{fig:lifescan} which shows that the fit results 
only become stable in all samples for a fit starting point 
larger than $1.0$~ps.

After all selection cuts already described,   
the size and composition of the \BP and \BZERO~enhanced samples 
are summarised in Table~\ref{tab:sample1}.
The  \BP(\BZERO) sample sizes correspond to
a selection efficiency, with respect to the starting number of \BP~and \BZERO~states in 
the hadronic sample,
of  10.1\%(3.8\%) for both 1994 and  1995 data.
{\normalsize
\begin{table}[h]
\begin{center}
    \begin{tabular}{|c|c|c|} \hline
                                               &  \BP~Sample        & \BZERO~Sample \\ \hline 
Real data sample size 1994(1995)               & 27356(13150) & 9293(4335) \\
Simulation sample size 1994(1995) $q\bar{q}$   & 61821(23415) & 22667(8533) \\
Simulation sample size 1994(1995) $b\bar{b}$   & 161198(42951)& 59015(15863)  \\
\BP~fraction                                   & $71.7 \%$        & $15.8 \%$          \\
\BZERO~fraction                                & $20.7 \%$        & $68.5 \%$          \\
\BS~fraction                                   & $3.7 \%$         &  $10.1 \%$         \\
b-baryon fraction                              & $2.4 \%$         &  $4.4 \%$           \\
u,d,s fraction                                 & $0.2 \%$         &  $0.1 \%$       \\
c fraction                                     & $1.2 \%$         &  $1.0 \%$       \\\hline
\end{tabular}
\caption[]
{\it\label{tab:sample1} The \BP~and \BZERO~sample size in data and 
simulation and the composition of the simulation. The simulation 
has been weighted for the quantities listed in Section~\ref{sec:weight}. } 
\end{center}
\end{table}
}

The data sample used to fit for the mean b-hadron lifetime
passed through the same  event selection and proper time
cuts as for the  \BP~and  \BZERO~samples but without any requirement
on the $NN(B_x)$~neural network outputs.
The size and composition of the mean b-hadron
 sample is summarised in Table~\ref{tab:sample2}. These numbers imply that
the mean lifetime measurement is valid for a b-hadron mixture, as given
by the simulation, of \BP$=41.9 \%$, \BZERO$=41.2 \%$, \BS$=8.9 \%$ and  
b-baryon$=8.0 \%$.

{\normalsize
\begin{table}[h]
\begin{center}
    \begin{tabular}{|c|c|} \hline
                          &  b-hadron    \\ \hline 
Sample size 1994(1995)    &  114317(54958)           \\
Simulation sample size 1994(1995) $q\bar{q}$   & 262697(98230) \\
Simulation sample size 1994(1995) $b\bar{b}$   & 677998(180634)    \\
\BP~fraction                &  $41.2 \%$                         \\
\BZERO~fraction             &  $40.5 \%$                         \\
\BS~fraction                &  $8.7 \%$                         \\
b-baryon fraction           &  $7.7 \%$                         \\
u,d,s fraction              &  $0.3 \%$                         \\
c fraction                  &  $1.7 \%$                         \\\hline
\end{tabular}
\caption[]
{\it\label{tab:sample2} The mean b-hadron sample size in data and 
simulation and the composition of the simulation. The simulation 
has been weighted for the quantities listed in Section~\ref{sec:weight}. } 
\end{center}
\end{table}
}

\subsection{Lifetime Results}
\label{sec:fit}
The \BP, \BZERO~and \meanb~lifetimes were extracted by fitting the 
simulated proper time distribution to the same distribution formed in the data
using a binned $\chi^2$ method. 
As discussed in Section~\ref{sec:workpt} the  start point of the 
fit range was chosen to be $1$~ps. The upper limit was positioned 
to avoid the worst effects of 
spurious, mainly two-track vertices, with very long reconstructed lifetimes 
while still accepting the vast majority of the data available. 
Nominally 100 bins were chosen but the exact binning was determined by the 
requirement that at least 10 entries be present in all bins of the data distribution.

To avoid the need to generate many separate Monte Carlo samples with 
different B-lifetimes, weighting factors
were formed for each lifetime measurement from the ratio
of exponential decay probability functions. 
Specifically, the weight,
\begin{eqnarray}
\nonumber
  w_i=\frac{\tau_{old}}{\tau_{new}}exp\left(\frac{t_i(\tau_{new}-\tau_{old})}{\tau_{old}\tau_{new}}\right),
\nonumber
\end{eqnarray}
for measurement $i$ and true B-lifetime $t_i$,  effectively transforms
the Monte Carlo lifetimes generated with a mean lifetime $\tau_{old}$ to be
distributed with a new mean value of $\tau_{new}$. Throughout the fit for the 
 \BP~and \BZERO~lifetimes, the ${\mathrm B_s}$~and ${\mathrm \Lambda_b}$~lifetime components 
were weighted to the current world average numbers listed in Table~\ref{tab:bfracs} and
for the \meanb~fit, the starting value in the simulation was $1.6\,$ps.   
The $\chi^2$~function given below was then minimised with respect to the 
\BP~and \BZERO~lifetimes in a simultaneous two parameter fit  
or to the mean b-hadron lifetime \meanb~in a one parameter fit,
\begin{eqnarray}
\nonumber
  \chi^2=\sum_{{\mathrm B}^0,{\mathrm B}^+}\left[\sum_{i=1}^{n_{bins}}\frac{(W^{sim}_i-N^{data}_i)^2}{(\sigma^{sim}_{i})^2+(\sigma^{data}_{i})^2}\right].
\nonumber
\end{eqnarray}
Here, $N^{data}_i$~is the number of data entries in bin $i$~and  $W^{sim}_i$~is  
the corresponding sum of weights in bin $i$~of the simulation. 

The results from  all lifetime fits, after imposing the working point conditions and 
following the above procedure,
are listed in Table~\ref{tab:fitresults}. In the table, the first error
quoted is statistical and the second systematic. The various 
sources of systematic error 
are described in Section~\ref{sec:syst}. Results are given 
for 1994 and 1995 data separately and combined taking into account
correlated systematic errors as described in Section~\ref{sec:conclude}.

\begin{figure}[htb]
\begin{center}
\epsfig{file=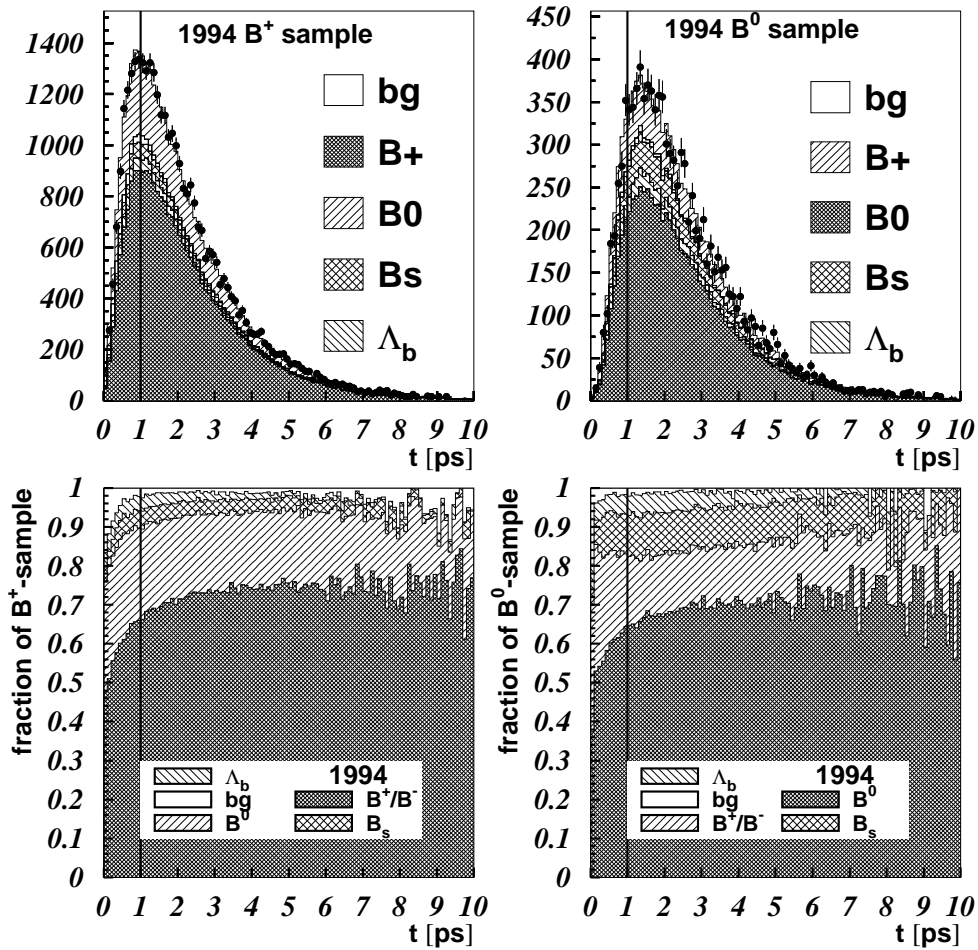,width=12.9cm}
\caption[]{\it\label{fig:fits} The upper two plots show the result of the fit in 
the \BP(left)~and 
\BZERO(right) samples in 1994(histogram) compared to data(points).  The b-hadron  
composition of the \BP~and \BZERO~sample is also indicated 
where `bg' refers to  
the background from non-${\mathrm b}\bar{{\mathrm b}}$~${\mathrm Z}^0$~decays.
The lower two plots trace how the fractional composition of the sample changes
in bins of the reconstructed lifetime.
The vertical line at $t=1\, \mathrm{ps}$~indicates that data below this point are removed
from the analysis.
}
\end{center}
\end{figure}

{\normalsize
\begin{table}[h]
\begin{center}
    \begin{tabular}{|c|c|c|c|} \hline
b-State & \multicolumn{3}{c|} {Fitted Lifetime}                                                  \\ \hline 
        & '94                    & '95                  & Combined        \\ \hline \hline     
\BP     & $1.624\pm 0.017\pm0.023$~ps& $1.623\pm 0.025\pm 0.019$~ps&$1.624\pm 0.014 \pm 0.018$~ps \\ \hline
\BZERO  & $1.548\pm 0.026\pm 0.035$~ps& $1.497\pm 0.039\pm 0.041$~ps& $1.531 \pm 0.021 \pm 0.031$~ps \\ \hline

$\frac{\tau_{{\mathrm B}^+}}{\tau_{{\mathrm B}^0}}$& $1.049 \pm 0.025 \pm 0.027$ & $1.085 \pm 0.040 \pm 0.036$ &
$1.060 \pm 0.021 \pm 0.024$ \\ \hline  
\meanb  & $1.577\pm 0.006\pm 0.008$~ps& $1.555\pm 0.009 \pm 0.011$~ps & $1.570 \pm 0.005 \pm 0.008$~ps \\ \hline
\end{tabular}
\caption[]
        {\it\label{tab:fitresults} The results of the lifetime fits
 in the 1994 and 1995  data samples where the first error quoted
is statistical and the second systematic. }
\end{center}
\end{table}
} 
The \BP~and \BZERO~fits (to the 1994 data) are shown in Figure~\ref{fig:fits}. 
The correlation coefficient between \BP~and \BZERO~lifetimes was found to be 
$-0.51$~for both the 1994 and 1995 fits.  
The fit 
$\chi^2$~at the minimum point was $178$~for $160$~degrees of freedom for 1994
data and $142$~for $143$~degrees of freedom for 1995 data.
 
\begin{figure}[h]
\begin{center}
\leavevmode
\epsfig{file=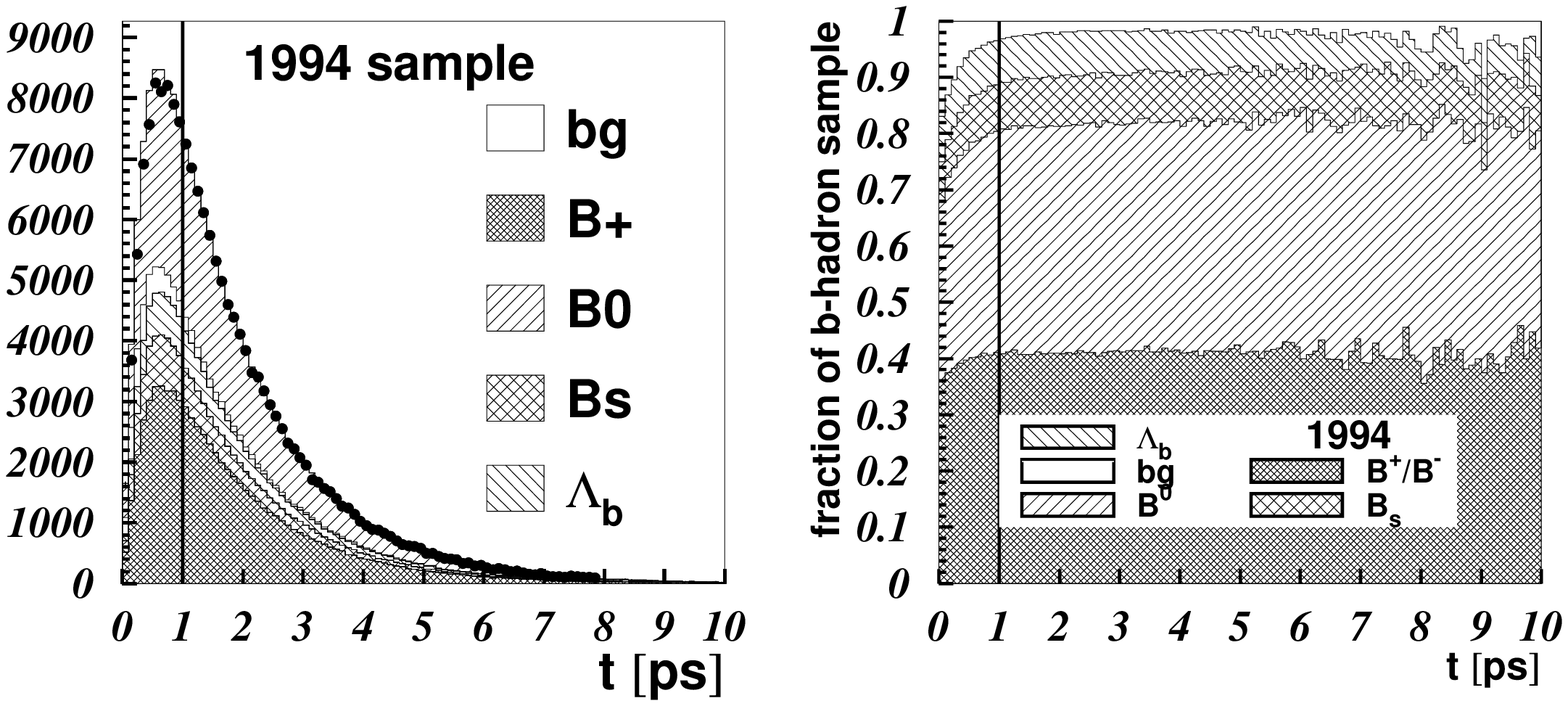,width=14.cm}
\caption[]
        {\it\label{fig:meanfit} The left plot shows the result of the mean
          $b$-hadron lifetime fit in 1994 (histogram) compared to the 
        data(points) at the working
        point. The right plot traces how the fractional  composition 
varies in bins of the reconstructed lifetime. 
The vertical line at $t=1\,\mathrm{ps}$~indicates that data below this point are removed
from the analysis.
}
\end{center}
\end{figure}

The mean b-hadron lifetime fit   
is shown in Figure~\ref{fig:meanfit}.
The $\chi^2$~at the minimum point was $76$~for $88$~degrees of 
freedom in 1994 data and $70$~for $88$~degrees of freedom for 1995 data.  

\begin{figure}[htb]
\begin{center}
\epsfig{file=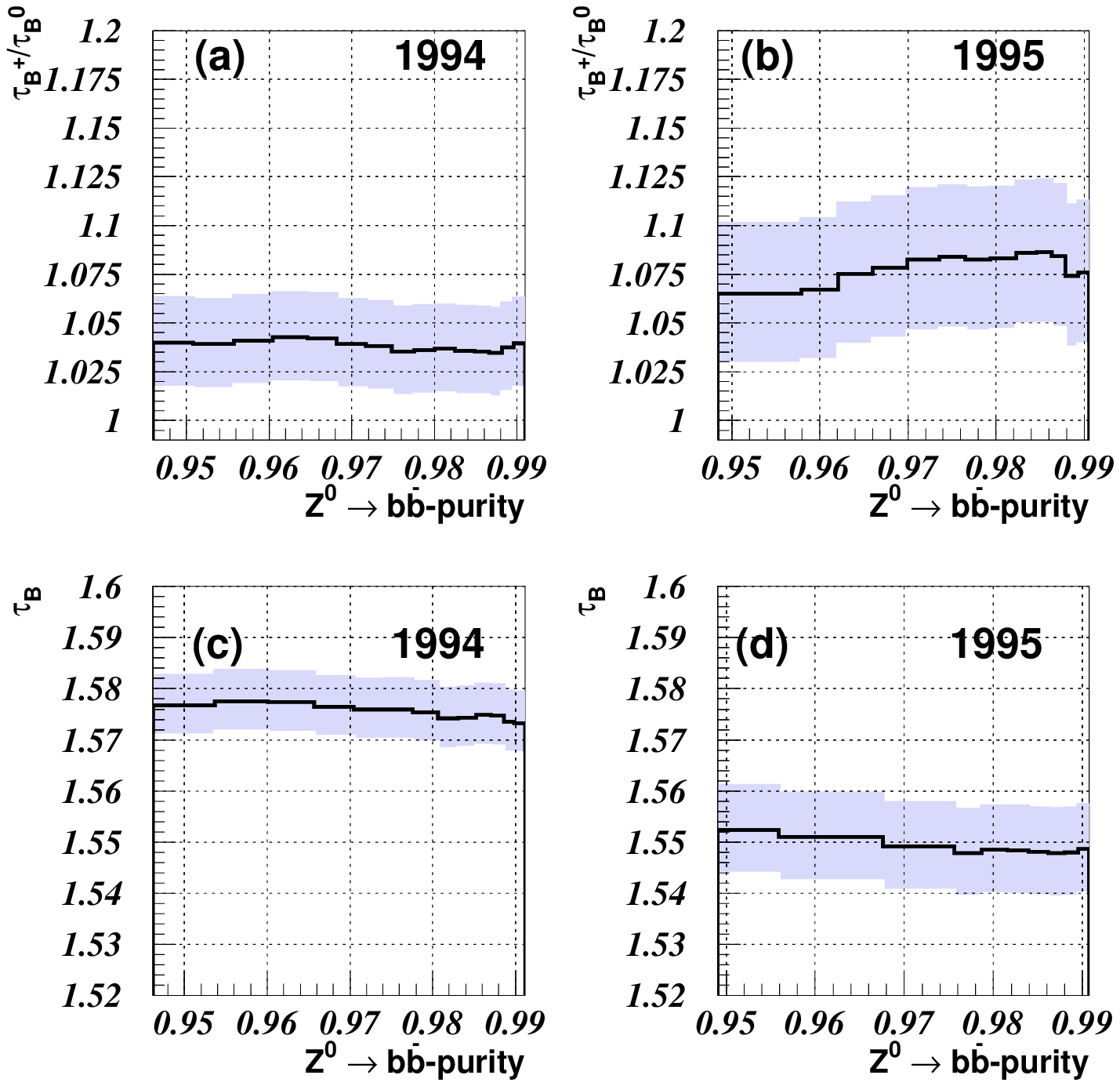,width=13.5cm}
\caption[]
        {\it\label{fig:stabil} The variation in the fitted lifetimes
as a function of the \Ztobb~purity for the ratio $\tau(B^+)/\tau(B^0)$~in
(a) 1994 and (b) 1995 data and for the mean b-hadron lifetime
in (c) 1994 and (d) 1995 data.
The upper and lower shaded bands represent the statistical one standard 
deviation errors which are correlated bin-to-bin.  } 
\end{center}
\end{figure}

\begin{figure}[p]
\begin{center}
\leavevmode
\epsfig{file=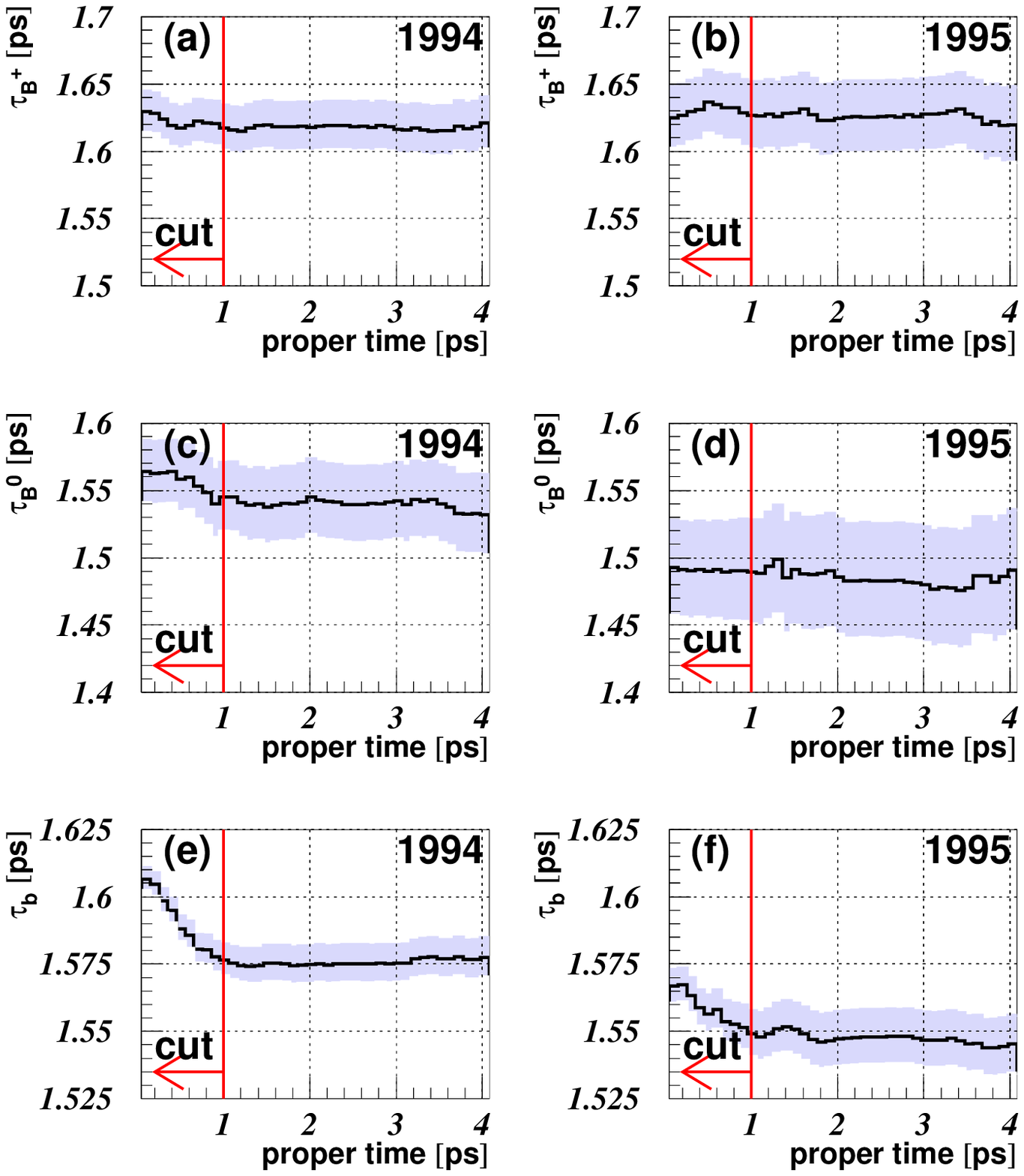,width=15.0cm}
\caption[]{\it\label{fig:lifescan} Lifetime fit results as a function of varying the start point of the fit
for \BP~in (a) 1994 and (b) 1995, \BZERO~in (c) 1994 and (d) 1995
and for the mean b-hadron fit in (e) 1994 and (f) 1995. }
\end{center}
\end{figure}

Figure~\ref{fig:stabil} illustrates the effect of cut scans
in the \Ztobb~event purity (i.e. the estimated fraction of the data sample
fitted coming from  \Ztobb~events)
showing a  good stability over a wide range 
of the cut values for the lifetime ratio $\tau(B^+)/\tau(B^0)$
and the mean b-hadron lifetime. Similarly Figure~\ref{fig:lifescan}
illustrates that the results are very stable over a wide range of different
start points for the fits above the default cut point of $t=1\,$ps.

A further crosscheck on the 
results was made by repeating the fits for one data set using the 
simulation sample compatible with  another data set 
e.g. fitting 1994 data using 1995 simulation.
It was found that all fit results
(for \BP, \BZERO~and \meanb) for both cases 
(1994 data using 1995 simulation and 1995 data using 1994 simulation)
changed by amounts that were within the systematic error for 
detector effects quoted in Table~\ref{tab:tab2} which  
provides a rough check that aspects of 
detector and physics modelling are well under control.

\section{Systematic Uncertainties}
\label{sec:syst}
Systematic uncertainties on the lifetime measurements come from 
three main sources. The first source is from the 
 modelling of heavy flavour physics parameters in our
Monte Carlo generator. Since attempts were made to model these effects using 
current world averages, these errors are largely irreducible. The second
source comes from the analysis method itself and the choices made in 
determining the measurement working point. The good level of
agreement between simulation and data and the fact that the result
is stable within a wide range of the working point 
(e.g. as shown in Figure~\ref{fig:stabil}) mean that these errors are kept to
a minimum. The third source of systematic uncertainty can be generically
termed `detector effects' and results from a less-than-perfect modelling
in simulation of the response of the detector.

Tables~\ref{tab:tab2}~and \ref{tab:tab3} present the 
full systematic error breakdown 
for the measurements of $\tau({\mathrm B}^+)$, $\tau({\mathrm B}^0)$~and
 \meanb~in 1994 and 1995 data.

\subsection{Heavy Flavour Physics Modelling}
\label{sec:systa}
Where possible, B-physics modelling uncertainties were estimated 
by varying central values by plus and minus one standard deviation  
and taking half of the observed change in the fitted lifetime 
value as the resulting systematic uncertainty from that source.

The \BS, b-baryon lifetimes,  b-hadron production fractions and 
fragmentation $\left< x \right>$~value 
have been varied within their errors as listed in 
Section~\ref{sec:weight} and 
half of the full variation in the results has been assigned as an error.
In the case of the b-hadron production fractions,
the variation was made taking into account correlations
from the 
covariance matrix listed in Table~\ref{tab:bfracs}.

{\normalsize
\begin{table}[h,p]
\begin{center}
\rotatebox{90}{\mbox{
\begin{tabular}{|l|c|c|c|c|c|c|c|} \hline
 \multicolumn{2}{|c|}{{}} &  \multicolumn{2}{|c|}{{\bf~~~$\tau_{{\mathrm B}^+}\,(1994 | 1995)$~~~}} &  \multicolumn{2}{|c|}{{\bf~~~$\tau_{{\mathrm B}^0}\,(1994 | 1995)$~~~}} &  \multicolumn{2}{|c|}{{\bf~~~$\frac{\tau_{{\mathrm B}^+}}{\tau_{{\mathrm B}^0}}\,(1994 | 1995)$~~~}}\\ \hline
\multicolumn{2}{|l|}{\bf Result [ps]}                            & {\bf 1.6241} &{\bf 1.6233}  & {\bf 1.5483} & {\bf 1.4971} & {\bf 1.0492} & {\bf 1.0848}\\ \hline
\multicolumn{2}{|l|}{\bf Statistical Error [ps]}                 & {\bf 0.0168} & {\bf 0.0251} &{\bf 0.0255}  &{\bf 0.0388} & {\bf 0.0247} & {\bf 0.0398}\\ \hline 
{\bf Source of Systematic Error} & {\bf~~~~~~Range~~~~~~} & \multicolumn{2}{|c|}{{\bf~~~$\Delta \tau_{{\mathrm B}^+}$~}[ps]} & \multicolumn{2}{|c|}{{\bf~~~$\Delta \tau_{{\mathrm B}^0}$~}[ps]} & \multicolumn{2}{|c|}{{\bf~~~$\Delta\frac{\tau_{{\mathrm B}^+}}{\tau_{{\mathrm B}^0}}$~}}\\ \hline
{Physics Modelling}& &\multicolumn{2}{|c|}{}&\multicolumn{2}{|c|}{}& \multicolumn{2}{|c|}{} \\ \hline 
 ${\mathrm B}_s$ lifetime        & $1.464\pm0.057$~ps    & \multicolumn{2}{|c|}{0.0007} & \multicolumn{2}{|c|}{0.0080}   & \multicolumn{2}{|c|}{0.0050}   \\  
 ${\mathrm b}$-baryon lifetime   & $1.208\pm0.051$~ps    & \multicolumn{2}{|c|}{0.0007}  & \multicolumn{2}{|c|}{0.0030}   & \multicolumn{2}{|c|}{0.0028} \\
 ${\mathrm b}$-hadron prod. fractions & See text & \multicolumn{2}{|c|}{0.0035} & \multicolumn{2}{|c|}{0.0035} & \multicolumn{2}{|c|}{0.0004} \\ 
 Fragmentation function    & $\left< x \right>=0.7153 \pm 0.0052$~\cite{delphi_bfrag}  & \multicolumn{2}{|c|}{0.0037} & \multicolumn{2}{|c|}{0.0026} & \multicolumn{2}{|c|}{0.0040} \\
${\mathrm B}\rightarrow{\mathrm D X}$ branching fractions & See text & \multicolumn{2}{|c|}{0.0081} & \multicolumn{2}{|c|}{0.0086} & \multicolumn{2}{|c|}{0.0017}\\
BR(${\mathrm B}\rightarrow$ wrong-sign charm) & $20.0\% \pm 3.3\%$ &  \multicolumn{2}{|c|}{0.0030} & \multicolumn{2}{|c|}{0.0047} & \multicolumn{2}{|c|}{0.0014} \\       
BR(${\mathrm B_s}\rightarrow {\mathrm D_s}$) & $35\% \rightarrow 70\%$  &  \multicolumn{2}{|c|}{0.0017} & \multicolumn{2}{|c|}{0.0076} & \multicolumn{2}{|c|}{0.0062} \\
BR$({\mathrm b} \rightarrow$c-baryon X) & $9.6\% \pm 3.0\%$ &   \multicolumn{2}{|c|}{0.0009} &   \multicolumn{2}{|c|}{0.0032} &   \multicolumn{2}{|c|}{0.0016} \\                
 ${\mathrm D^+},{\mathrm D^0}$ topo. branching ratios &  \cite{DBRS} & \multicolumn{2}{|c|}{0.0016}  & \multicolumn{2}{|c|}{0.0113} & \multicolumn{2}{|c|}{0.0084} \\
B meson mass &  $m_B=5.2789\pm0.0018~\mathrm{Gev/c^2}$  & \multicolumn{2}{|c|}{ 0.0004} & \multicolumn{2}{|c|}{ 0.0015} & \multicolumn{2}{|c|}{0.0011}  \\
\hline
{b-hadron Reconstruction}& &\multicolumn{2}{|c|}{}&\multicolumn{2}{|c|}{}& \multicolumn{2}{|c|}{} \\ \hline 
   b and c efficiency correction & On/off                             & 0.0076& 0.0048 & 0.0050& 0.0066 & 0.0015& 0.0017 \\           
    $NN(B_x)$~cuts        & 65\%-75\% purity                     & 0.0093 & 0.0133 & 0.0216 & 0.0255 & 0.0196 & 0.0269 \\
    $NN(B_x)$~shape       & See text                             & 0.0008 & 0.0006 & 0.0099 & 0.0126 & 0.0059 & 0.0097\\
   Sec. vertex multiplicity & See text   & 0.0022 & 0.0019 & 0.0042 & 0.0040 & 0.0040 & 0.0042  \\  
\hline                          
{Detector Effects}& &\multicolumn{2}{|c|}{}&\multicolumn{2}{|c|}{}& \multicolumn{2}{|c|}{} \\ \hline
   Resolution and     &   On/off         & 0.0171 & 0.0075& 0.0144& 0.0200 & 0.0110 & 0.0159 \\
   hemisphere quality &                  &        &       &       &        &        & \\ \hline
    {\bf Total Systematic Error}&                & {\bf 0.0234} & {\bf 0.0192} & {\bf 0.0345} & {\bf 0.0406} &{\bf 0.0269}& {\bf 0.0355} \\ \hline 
\end{tabular}
}}
\caption[] {\label{tab:tab2}\it Summary of systematic
  uncertainties in the \BP and \BZERO~lifetime results and their ratio for 1994 and 1995 data. 
Systematic errors are assumed independent and added in quadrature to give
 the total systematic  error quoted.} 
 \end{center}
\end{table}
}

{\normalsize
\begin{table}[h]
\begin{center}
  \hspace*{-0.9cm}
    \begin{tabular}{|l|c|c|c|}
  \hline
 \multicolumn{2}{|c|}{} & {\bf~~~$\tau_{{\mathrm B}}$ for 1994~~~} & {\bf~~~$\tau_{{\mathrm B}}$ for 1995~~~} \\ \hline
\multicolumn{2}{|l|}{\bf Result [ps]}                             & {\bf 1.5773} & {\bf 1.5546} \\ \hline
\multicolumn{2}{|l|}{\bf Statistical Error [ps]}                 & {\bf 0.0059} & {\bf 0.0085} \\ \hline 
{\bf Source of Systematic Error} & {\bf~~~~~~Range~~~~~~} & {\bf~~~$\Delta \tau_{{\mathrm B}}$~[ps]} & {\bf~$\Delta \tau_{{\mathrm B}}$~[ps]} \\ \hline
Physics Modelling& & &  \\ \hline 
  ${\mathrm b}$-hadron prod. fractions & See text                                        & \multicolumn{2}{|c|}{0.0011}  \\ 
 Fragmentation function    & $\left< x \right>=0.7153 \pm 0.0052$~\cite{delphi_bfrag}                         & \multicolumn{2}{|c|}{0.0038}  \\
BR(${\mathrm B}\rightarrow$ wrong-sign charm) & $20\% \pm 3.3\%$                           & \multicolumn{2}{|c|}{0.0010} \\       
BR(${\mathrm B_s}\rightarrow {\mathrm D_s}$) & $35\%\rightarrow 70\%$                    & \multicolumn{2}{|c|}{0.0012}   \\        
BR$({\mathrm b} \rightarrow$c-baryon X) & $9.6\% \pm 3.0\%$                     & \multicolumn{2}{|c|}{0.0017}   \\        
${\mathrm B}\rightarrow{\mathrm D X}$ branching fractions & See text                       & \multicolumn{2}{|c|}{0.0008} \\
${\mathrm D^+},{\mathrm D^0}$ topo. branching ratios &  See text                      & \multicolumn{2}{|c|}{0.0008}  \\
B-meson mass &  $m_B=5.2789\pm0.0018~\mathrm{Gev/c^2}$                                                & \multicolumn{2}{|c|}{0.0004}  \\
\hline
 b-hadron Reconstruction  & & & \\ \hline
   b and c efficiency correction & On/off                                                & 0.0057 & 0.0048 \\           
   Sec. vertex multiplicity & See text                                                   & 0.0019 & 0.0023   \\  \hline
Detector Effects & & & \\ \hline
   Resolution and           & See text                                                   & 0.0035 & 0.0083 \\          
   hemisphere quality & & & \\ \hline
    {\bf Total Systematic Error}&                                                        & {\bf 0.0084} & {\bf 0.0109} \\ \hline
\end{tabular}
\caption[] {\label{tab:tab3}\it Summary of systematic
  uncertainties in the mean b-hadron~lifetime for 1994 and 1995 data. 
Systematic errors are assumed independent and added in quadrature to give
 the  total systematic error.} 
    \end{center}
\end{table}    
}
 
Close attention was paid to possible systematic effects 
on the analysis due to the modelling of  
${\mathrm b} \rightarrow $~charm branching ratios, where 
current experimental knowledge is scarce. The charm content 
impacts on the performance of the 
${\mathrm B}^+$~and ${\mathrm B^0}$~enhancement networks and 
can pull the reconstructed B-vertex position to longer decay 
lengths. The size of this pull in turn depends on whether a 
${\mathrm D}^0$~or ${\mathrm D}^+$~was produced since 
$\tau({\mathrm D}^+) \sim 2.5$~times larger than $\tau({\mathrm D}^0)$.  
Specific aspects of the Monte Carlo that were found to  
warrant systematic error contributions were:
\newline
(a) Inclusive, ${\mathrm B} \rightarrow {\mathrm D}$~branching ratios
were adjusted in the Monte Carlo according to a fit using all currently
available measurements from~\cite{PDG} as constraints.
The values taken were:
$BR\left( \bar{{\mathrm B}}^0 \rightarrow {\mathrm D}^+ X \right)=15.6\%$, 
$BR \left(\bar{\mathrm{B}}^0 \rightarrow {\mathrm D}^0 X \right)=65.8\%$,
$BR \left({\mathrm B}^- \rightarrow {\mathrm D}^+ X \right)=29.3\%$,
$BR \left({\mathrm B}^- \rightarrow {\mathrm D}^0 X \right)=52.1\%$.
The full difference seen in the results due to this change was assigned as
a systematic error.
\newline
(b) The standard Monte Carlo data set used 
contained a wrong-sign charm production rate into ${\mathrm D_s}$~of $11\%$.
This rate is now known to be too low due to the 
production of ${\mathrm D}^0/{\mathrm D}^+$~mesons at the 
$W$~vertex (in addition to  ${\mathrm D_s}$), and
an estimate  of the overall rate based on~\cite{HFWG} 
is BR(wrong-sign)$=20.0\% \pm 3.3 \%$.  
To account for this discrepancy with current  measurements,
the wrong-sign rate in the simulation was weighted up
to $20.0 \%$~and all quoted lifetime results 
were shifted to be valid for this higher wrong-sign rate.
An error was then assigned based on the change in the lifetime
results observed when the   wrong-sign rate was further changed
by $\pm 3.3\%$. The impact of the simulation containing
only wrong-sign ${\mathrm D_s}$~mesons instead of a mixture
of ${\mathrm D_s}, {\mathrm D}^0$~and ${\mathrm D}^+$~was 
tested with specially generated Monte Carlo data sets, and the 
effect found to be small compared to the overall effect of
having almost double the rate of events containing two D-mesons
per hemisphere.
\newline
(c) BR$({\mathrm B_s} \rightarrow {\mathrm D_s}$X) is currently known to,
at best, $\pm 30$\%~\cite{PDG} and was varied in the Monte Carlo
by a factor two from the default value of $35\%$. The full change 
in the fitted lifetime was then assigned as a systematic error.
\newline
(d) BR$({\mathrm b} \rightarrow$c-baryon X), where b~represents 
the natural mixture of b-mesons and baryons at LEP, was varied
from the default value in the simulation of 
$9.6 \%$~by $\pm 3.0 \%$. This range covers the  
uncertainty on this quantity from experiment which currently
stands at
BR$({\mathrm b} \rightarrow$c-baryon X)$=9.7 \pm 2.9 \%$~\cite{PDG}.
Half of the full change 
in the fitted lifetime was then assigned as a systematic error.  

The uncertainty from D-topological 
branching fractions was estimated from the difference
in the fit result obtained when weighting according to the results
from~\cite{DBRS}. 
The masses of B-mesons were varied within plus and minus 
one standard deviation 
of the value assumed in the BSAURUS package and half of the
change seen taken as a systematic error. 

Since many of the physics modelling systematics investigated are significantly
smaller than the statistical precision, the approach was taken to
average the errors, evaluated in  1994 and 1995 data, weighted by the 
statistical error for each year. This ensures that the effects
of statistical fluctuations in the determination of these errors
are minimised and explains why 
the physics modelling errors appearing in Table~\ref{tab:tab2} are 
the same for 1994 and 1995.

\subsection{b-hadron Reconstruction}
\label{sec:systb}
The efficiency for reconstructing ${\mathrm b}\bar{{\mathrm b}}$~and 
${\mathrm c}\bar{{\mathrm c}}$~events
(as a function of the event b-tag) has been extracted from the
real data by a double-hemisphere tagging technique.
At the b-tag value of the working point, 
the results of this study suggest that while the 
reconstruction efficiency for  ${\mathrm b}\bar{{\mathrm b}}$~events
might  be underestimated in the simulation by about $\sim 2 \%$~(relative), the 
 efficiency for  ${\mathrm c}\bar{{\mathrm c}}$~events in
simulation is $\sim 8 \%$~(relative) lower than in data. 
To account for this possible source of error 
the difference seen in the fit results, when these efficiencies were
changed in the simulation to agree with the numbers above, was
assigned as a systematic error. Since a large part
of the discrepancy between simulation and data in the  
${\mathrm c}\bar{{\mathrm c}}$~event reconstruction efficiency is 
due to a less-than-perfect  modelling of charm physics, this 
error contribution has already been partially accounted for by the 
explicit charm physics systematics detailed above. Given the 
current level of uncertainty in this sector,  the 
conservative approach of quoting both error contributions is preferred.


As was remarked in Section~\ref{sec:fit},
uncertainties resulting from the method itself have been checked by 
scanning regions around critical cut values to check for stability
as illustrated in  Figure~\ref{fig:stabil}.
 In addition, the 
binning used for the $\chi^2$~formulation was varied over a wide range 
as was the   minimum number of entries per bin (set by default to 10)
and were both  found to give  no significant change in the results.

The impact on the analysis of any residual discrepancy  
between data and simulation in the BD-Net
variable was checked by studying the effect of removing the
Strip-Down vertex algorithm (see Section~\ref{sec:decl})
from the analysis. Of the four vertex algorithms used, the 
Strip-Down method is the most sensitive to details of the BD-Net 
variable since it imposes a direct cut in the BD-Net
as part of the track selection.
Removing the Strip-Down algorithm and replacing it with one of the 
other three methods, selected by the same criteria    
as described in Section~\ref{sec:decl}, resulted in lifetime results
for 1994 data that changed by $\Delta \tau(B^+)=-0.0017$~ps and 
$\Delta \tau(B^0)=-0.0063$~ps i.e.  well within the
total systematic error quoted. In addition it was confirmed that 
the proper time distributions were well compatible
when the cut imposed on the  BD-Net distribution
was changed from the default value of zero to $\pm 0.2$.

The analysis assumes that the \BP~and \BZERO~purities are well modelled 
by the simulation.
A systematic error will arise if the shape of the 
$NN(B^+)$~and $NN(B^0)$~network outputs differ from the data
and/or the composition of the \BP~and \BZERO~simulated samples differ.
Any difference in shape between data and simulation 
was accounted for in the following way. 
It was assumed that the difference could be wholly accounted for by a 
change in just the \BP~composition for the case of the $NN(B^+)$~network
and  by just the \BZERO~composition for the case of the $NN(B^0)$~network.
In this way it was found that the maximum error
made by assuming that the  \BP~and  \BZERO~purities in the simulation were correct, 
was of order 2\% and 4\%, respectively. The effect of these changes were 
then propagated into errors on the extracted lifetimes.
To account for any composition differences, half of the maximum variation 
in the fitted lifetime while
scanning the purity range $\left[ 65\%,75\% \right]$~was assigned as an error. 
The scan range was chosen to enclose the largest uncertainty 
on the purity found from analysing the shapes of the network outputs
described above.

The multiplicity of tracks in the reconstructed b-hadron vertex was found to be
in overall good agreement between the data and simulation. To account for
any residual differences a weight was formed from the ratio of the data
and simulation distributions and the change seen as a result of applying this 
weight was assigned as a systematic error.


\subsection{Detector Modelling}
\label{sec:systc}
In order to account for uncertainties in the simulation originating from
detector response modelling, the effect
of switching on and off the  following corrections  was studied:
\begin{itemize}
\item the hemisphere quality weight, described in Section~\ref{sec:weight},
\item an attempt to obtain a better match of the track impact parameter and error (with
respect to the primary vertex) between simulation and data according
to the prescription detailed in \cite{AABTAG}.
\end{itemize}
Since in general, knowledge of detector modelling 
uncertainties are not at the same level of understanding as e.g. our knowledge
of the difference between the  B-production fractions in our Monte Carlo 
and the world averages, the following approach was taken to assigning systematic values
for these effects:
all four combinations of switching these corrections on/off for the 
\BP/\BZERO~fit and the \meanb~analysis
were made and the fitted lifetimes of the four possibilities recorded.
The central results for \BP,\BZERO~and \meanb~were then chosen 
to be the mean values of these four combinations, and 
the resulting systematic error from detector response modelling
was assigned to be
half of the maximum spread of the values recorded
from the four combinations. This error is listed in 
tables~\ref{tab:tab2}~and \ref{tab:tab3} as
`Resolution and hemisphere quality'.

\subsection{Closing Remarks on Systematic Errors}
 In general it can be concluded
 from Tables~\ref{tab:tab2}~and \ref{tab:tab3}
that detector
effects dominate. Physics modelling errors come essentially 
only from b-physics sources, since the contamination from 
light-quark and charm events is so small,
and are generally well under control. 
For the case of the \BP~and 
\BZERO, additional  systematic error contributions 
arising from the enhancement neural networks ( $NN(B_x)$)  
reflect the difficult task of modelling accurately such complex variables.

\vspace{-0.5cm} 
\section{Summary and Conclusion}
\label{sec:conclude}
The lifetimes of 
\BP, \BZERO, their ratio and the mean b-hadron lifetime 
have been measured. The analysis
isolated b-hadron candidates with neural network techniques trained to exploit the
physical properties of inclusive b-hadron decays. 
Binned $\chi^2$ fits to the resulting 
DELPHI data samples collected in 1994 and 1995 yielded
the results  presented in Table~\ref{tab:tab2}
for \BP~and \BZERO~and the result for the mean b-hadron lifetime
is presented in Table~\ref{tab:tab3}.
The results for 1994 and 1995 were combined, 
treating all systematic contributions as independent 
apart from the following (which were assumed to be $100 \%$~correlated):
\begin{itemize}
\item all physics modelling errors,
\item the  $NN(B_x)$~shape error,
\item the secondary vertex multiplicity error. 
\end{itemize}
The combined results for the \BP~and \BZERO~were,
\begin{tabbing}
tttttt\=ttt\=tttttt\=tttttttttttttttt\=ttttttttttttt \kill
\hspace{2.5cm}
$\tau_{{\mathrm B}^+}$ \>\hspace{2.5cm} = \>\hspace{2.5cm} 1.624\>\hspace{2.5cm} $\pm 0.014$~(stat)  \>\hspace{2.5cm}$\pm 0.018 $~(syst)~ps \\  

\hspace{2.65cm}$\tau_{{\mathrm B}^0}$ \>\hspace{2.5cm} = \>\hspace{2.5cm} 1.531 \>\hspace{2.5cm} $\pm 0.021$~(stat)  \>\hspace{2.5cm}$\pm 0.031 $~(syst)~ps \\

\hspace{2.45cm} $\frac{\tau_{{\mathrm B}^+}}{\tau_{{\mathrm B}^0}}$ \>\hspace{2.5cm} = \> \hspace{2.5cm} 1.060 \>\hspace{2.6cm}$\pm 0.021$~(stat)\> \hspace{2.5cm}$\pm 0.024$~(syst)
\end{tabbing}
and for the average b-hadron lifetime:
\begin{tabbing}
tttttt\=ttt\=tttttt\=tttttttttttttttt\=ttttttttttttt \kill
\hspace{2.65cm}$\tau_{{\mathrm b}}$ \>\hspace{2.5cm} = \>\hspace{2.5cm} 1.570 \>\hspace{2.5cm} $\pm 0.005$~(stat)  \>\hspace{2.5cm}$\pm 0.008$~(syst)~ps.
\end{tabbing}
(Note that the average b-hadron lifetime result is valid for a b-hadron mixture, as given
by the simulation, of \BP$=41.9 \%$, \BZERO$=41.2 \%$, \BS$=8.9 \%$ and  
b-baryon$=8.0 \%$).

These results are well compatible with previous DELPHI results in this area using 
the 1991-1993 data sets: based on ${\rm D}^{(*)} \ell$~reconstruction~\cite{DELPHI_life1} 
and~\cite{DELPHI_life2} and from an inclusive secondary vertex 
approach~\cite{DELPHI_TOPO}. No attempt has been made to combine these older results
with the current analysis because of the vast difference in precision e.g.
the error on the lifetime ratio  $\frac{\tau_{{\mathrm B}^+}}{\tau_{{\mathrm B}^0}}$~is
now a factor five better than was achieved in~\cite{DELPHI_TOPO}.

Compared to existing measurements, the \BP~lifetime result is currently the most accurate
and is well compatible with all other measurements and 
with the world average value of $1.671 \pm 0.018$~ps~\cite{PDG}.
The precision of the \BZERO~lifetime result is similar to 
the best so far achieved from ${\mathrm Z}^0$~data i.e. from an 
OPAL analysis based on inclusive ${\rm D}^{(*)}\ell$~reconstruction
($1.541 \pm 0.028 \pm 0.023$~ps~\cite{OPAL_B0}) and to recent results
from the B-factory experiments, BABAR~\cite{BABAR2} and BELLE~\cite{BELLE}. 
In addition, the  \BZERO~lifetime result is well compatible with all other
measurements and with the current world average value of $1.537 \pm 0.015$~ps~\cite{PDG}.
All published measurements of the lifetime ratio  $\frac{\tau_{{\mathrm B}^+}}{\tau_{{\mathrm B}^0}}$~are presented in Table~\ref{TAB:ALLRATIO} together with their
average~\cite{PDG}.
\begin{table}[bth]
\begin{center}
\begin{tabular}{|l|c|c|c|c|} \hline
 Experiment    & Method            & Data set      & Ratio $\tau_{{\rm B}^+} /\tau_{{\rm B}^0}$            & Reference \\   \hline
ALEPH          & D$^{(*)} \ell$    &  91-95        & 1.085$\pm 0.059 \pm 0.018$        & \cite{ALEB01} \\
ALEPH          & Exclusive rec.    &  91-94         & 1.27$^{+0.23+0.03}_{-0.19-0.02}$  & \cite{ALEB0}  \\
CDF            & D$^{(*)} \ell$   & 92-95         & 1.110$\pm 0.056^{+0.033}_{-0.030}$  & \cite{CDFB02}\\
CDF            & Excl. ($J/\psi K$)& 92-95         & 1.093 $\pm$ 0.066 $\pm$ 0.028          & \cite{CDFB01}\\
DELPHI         & D$^{(*)} \ell$    &  91-93        & 1.00$^{+0.17}_{-0.15} \pm$ 0.10     & \cite{DELPHI_life1}\\
DELPHI         & Charge sec. vtx.  &  91-93        & 1.06$^{+0.13}_{-0.11} \pm 0.10$     & \cite{DELPHI_TOPO}\\
L3             & Charge sec. vtx.  &  94-95        & 1.09$\pm$  0.07  $\pm$ 0.03         & \cite{L3B01} \\ 
OPAL           & D$^{(*)} \ell$    &  91-93        & 0.99$ \pm 0.14^{+0.05}_{-0.04}$     & \cite{OPAB0} \\ 
OPAL           & Charge sec. vtx.  &  93-95        & 1.079$\pm$ 0.064  $\pm$ 0.041       & \cite{OPAB1} \\ 
SLD            & Charge sec. vtx. $\ell$&  93-95   & 1.03$^{+0.16}_{-0.14} \pm$ 0.09     & \cite{SLDB01} \\ 
SLD            & Charge sec. vtx.  &  93-95        & 1.01$^{+0.09}_{-0.08} \pm$ 0.05 & \cite{SLDB01} \\
BABAR          & Exclusive rec.    &  99-00        & 1.082 $\pm$ 0.026$\pm$ 0.012        & \cite{BABAR1} \\
BELLE          & Exclusive rec.    &  99-01        & 1.091 $\pm$ 0.023$\pm$ 0.014        & \cite{BELLE} \\\hline
Average        &                   &               & 1.085 $\pm$ 0.017                   &    \\   \hline

\end{tabular}
\caption[]{Measurements of the ratio $\tau_{{\rm B}^+} /\tau_{{\rm B}^0}$.}
\label{TAB:ALLRATIO}
\end{center}
\end{table}

It can be seen that the result from this analysis is currently the most precise of  the 
measurements from ${\mathrm Z}^0$~decay data and the CDF experiment at the Tevatron and also 
has a precision comparable to the B-factory experiments BABAR and BELLE. Within the quoted errors
the result is also compatible with all measurements and with the world average value.


The result for the mean b-hadron lifetime significantly improves on the
most precise existing measurement from L3,  
$\tau_{{\mathrm b}}=1.556 \pm 0.010 \pm 0.017$~ps~\cite{L3_MEANT}
and is in good agreement with the most precise previous 
DELPHI publication on this subject~\cite{DELPHI_MEANT}.
In addition it is compatible with the current world average,
$1.564 \pm 0.014$~ps~\cite{PDG}, which has been compiled assuming 
that all measurements are based on b-hadron samples
with the same mixture of b-hadron species i.e. the b-hadron production 
fractions from $\rm{Z^0}$~decay. It is also informative to compare these  
values with the inclusive lifetime defined as, \\
$\tau_{\rm b} = \sum_i f({\rm B}_i) \tau({\rm B}_i)$, calculated using the current
world average values for b-hadron production  fractions $f({\rm B}_i)$~and 
lifetimes $\tau({\rm B}_i)$~from~\cite{PDG}: $\tau_{\rm b} = 1.543$~ps.

\subsection*{Acknowledgements}
\vskip 3 mm
 We are greatly indebted to our technical 
collaborators, to the members of the CERN-SL Division for the excellent 
performance of the LEP collider, and to the funding agencies for their
support in building and operating the DELPHI detector.\\
We acknowledge in particular the support of \\
Austrian Federal Ministry of Education, Science and Culture,
GZ 616.364/2-III/2a/98, \\
FNRS--FWO, Flanders Institute to encourage scientific and technological 
research in the industry (IWT), Belgium,  \\
FINEP, CNPq, CAPES, FUJB and FAPERJ, Brazil, \\
Czech Ministry of Industry and Trade, GA CR 202/99/1362,\\
Commission of the European Communities (DG XII), \\
Direction des Sciences de la Mati$\grave{\mbox{\rm e}}$re, CEA, France, \\
Bundesministerium f$\ddot{\mbox{\rm u}}$r Bildung, Wissenschaft, Forschung 
und Technologie, Germany,\\
General Secretariat for Research and Technology, Greece, \\
National Science Foundation (NWO) and Foundation for Research on Matter (FOM),
The Netherlands, \\
Norwegian Research Council,  \\
State Committee for Scientific Research, Poland, SPUB-M/CERN/PO3/DZ296/2000,
SPUB-M/CERN/PO3/DZ297/2000, 2P03B 104 19 and 2P03B 69 23(2002-2004)\\
JNICT--Junta Nacional de Investiga\c{c}\~{a}o Cient\'{\i}fica 
e Tecnol$\acute{\mbox{\rm o}}$gica, Portugal, \\
Vedecka grantova agentura MS SR, Slovakia, Nr. 95/5195/134, \\
Ministry of Science and Technology of the Republic of Slovenia, \\
CICYT, Spain, AEN99-0950 and AEN99-0761,  \\
The Swedish Natural Science Research Council,      \\
Particle Physics and Astronomy Research Council, UK, \\
Department of Energy, USA, DE-FG02-01ER41155. 



\end{document}